\documentclass[onecolumn,pre,floats,aps,amsmath,amssymb,nofootinbib]{revtex4-2}
\usepackage{comment}
\usepackage[utf8]{inputenc}
\usepackage{mathtools}
\usepackage{bbm}
\usepackage{cancel} 
\usepackage{dcolumn}
\usepackage{blindtext}
\usepackage{xcolor}
\usepackage{amsthm}
\usepackage{bm}
\usepackage{hyperref}
\usepackage{graphicx}
\usepackage{graphics,epsfig}
\usepackage{subcaption}
\usepackage{verbatim}
\usepackage{microtype}
\usepackage{adjustbox}
\usepackage{slashed}
\usepackage{physics}
\newcommand{\be}{\begin{equation}}
\newcommand{\ee}{\end{equation}}
\newcommand{\bea}{\begin{eqnarray}}
\newcommand{\eea}{\end{eqnarray}}
\newcommand{\bs}{\boldsymbol}

\newcommand{\beq}{\begin{equation}}
\newcommand{\eeq}{\end{equation}}

\newcommand{\nn}{\nonumber}

\newcommand{\Romatre}{Dipartimento di Matematica e Fisica, Universit\`a  Roma Tre and INFN, Sezione di Roma Tre,\\ Via della Vasca Navale 84, I-00146 Rome, Italy}
\newcommand{\RomatreINFN}{Istituto Nazionale di Fisica Nucleare, Sezione di Roma Tre,\\ Via della Vasca Navale 84, I-00146 Rome, Italy}
\newcommand{\Romadue}{Dipartimento di Fisica and INFN, Universit\`a di Roma ``Tor Vergata",\\ Via della Ricerca Scientifica 1, I-00133 Roma, Italy}

\begin{document}
\title{Spectral-function determination of complex electroweak amplitudes with lattice QCD} 
\author{R.\,Frezzotti}\affiliation{\Romadue} 
\author{G.\,Gagliardi}\affiliation{\RomatreINFN}
\author{V.\,Lubicz}\affiliation{\Romatre} 
\author{F.\,Sanfilippo}\affiliation{\RomatreINFN}
\author{S.\,Simula}\affiliation{\RomatreINFN}
\author{N.\,Tantalo}\affiliation{\Romadue}

\date{\today}

\begin{abstract}

We present a novel method to determine on the lattice both the real and imaginary parts of complex electroweak amplitudes involving two external currents and a single hadron or the QCD vacuum in the external states. The method is based on the spectral representation of the relevant time-dependent correlation functions and, by extending the range of applicability of other recent proposals built on the same techniques, overcomes the difficulties related to the analytic continuation from Minkowskian to Euclidean time, arising when intermediate states with energies smaller than the external states contribute to the amplitude. In its simplest form, the method relies on the standard $i \varepsilon$ prescription to regularize the Feynman integrals and at finite $\varepsilon$ it requires to verify the condition $1/L \ll \varepsilon \ll \Delta(E)$, where $L$ is the spatial extent of the lattice and, for any given energy $E$, $\Delta(E)$ represents the typical size of the interval around $E$ in which the hadronic amplitude is significantly varying. In order to illustrate the effectiveness of this approach in a realistic case, we apply the method to evaluate non-perturbatively the hadronic amplitude contributing to the radiative leptonic decay $D_s \to\ell\nu_\ell\,\gamma^*$, working for simplicity with a single lattice ensemble at fixed volume and lattice spacing.

\end{abstract}

\maketitle

\vspace{0.2cm}

\section{Introduction}
In the lattice regularization of QCD, the Wick rotation from real to imaginary time is a necessary ingredient to perform numerical simulations based on Montecarlo algorithms. As well known, this rotation introduces theoretical and numerical difficulties. A class of problems occurs in the study of processes with more than one hadron in the initial or final state, as firstly pointed out in the seminal paper by Maiani and Testa~\cite{Maiani:1990ca}. However, even when only a single hadron or the QCD vacuum appear in the external states, the analytic continuation from Minkowskian to Euclidean time may still be problematic. This occurs in particular in electro-weak amplitudes when one or more of the intermediate states contributing to a given process have energies smaller than those of the external states. In such a situation, the physical (Minkowskian) amplitude develops an imaginary part, and the integral over Euclidean time, which is usually considered on the lattice in order to evaluate the hadronic amplitude, diverges in the limit of infinite temporal extension $T$ of the lattice

In Ref.~\cite{Barata:1990rn} the problem of overcoming the difficulties induced by the Wick rotations in numerical simulations of lattice QCD has been solved, on the theoretical side, by relying on the spectral representation of the Euclidean lattice correlation functions and the theoretical connection between a generic amplitude and Euclidean lattice correlators has been established. More recently, this connection has been worked out in details in Ref.~\cite{Bulava:2019kbi} by considering  spectral densities smeared in energy with Cauchy kernels, depending upon the smearing parameter $\varepsilon$, which in the limit $\varepsilon\mapsto 0^+$ implement the standard $i 0^+$-regularization of Feynman integrals. 
In Ref.\,\cite{Bruno:2020kyl}, it is has been suggested that the limitations imposed by the Maiani-Testa theorem can also be circumvented by introducing in the Euclidean time integrals properly smeared theta functions, that allow to select the desired energy region even in the large time limit of the lattice correlation functions.

In this paper, we generalize these spectral densities techniques to overcome the problem of the analytic continuation from Minkowskian to Euclidean time arising in processes containing a single hadron or the QCD vacuum in the external states. Our proposal relies on the spectral representation of the hadronic amplitude and on the HLT method of Ref.\,\cite{Hansen:2019idp} for evaluating the spectral function smeared with a proper kernel. In its simplest form, as in the case of Ref.~\cite{Bulava:2019kbi}, the method proposed here relies on the standard $i \varepsilon$ prescription to regularize the Feynman integrals, which allows to write a generic hadronic amplitude $H(E)$, which is a function of the energy $E$, in terms of its spectral representation in the form
\be
\label{master0}
H(E) = \lim_{\varepsilon \to 0} \ \int_{E^*}^{\infty} \frac{dE'}{2\pi} \, \frac{\rho(E')}{E'-E - i \varepsilon} \ ,
\ee
where $\rho(E)$ is the spectral density and the lower integration limit $E^*$ is such that $\rho(E) =0$ for $E<E^*$. Since the integral in the spectral representation of Eq.\,\eqref{master0} is independent of time, all the difficulties associated with the analytic continuation from Minkowskian to Euclidean time are thus circumvented. 

A central ingredient of the proposal is the observation~\cite{Bulava:2019kbi,Hansen:2019idp,Bulava:2021fre} that, for finite values of $\varepsilon$, the integral of Eq.\,\eqref{master0} can be computed using the HLT method of Ref.\,\cite{Hansen:2019idp}, starting from the lattice determination of the relevant time-dependent correlation function. Therefore, by performing the calculations at different values of $\varepsilon$ and then extrapolating the results to $\varepsilon \to 0$, one can achieve a lattice determination of both the real and imaginary part of the hadronic amplitude. 

Two observations are now in order. The first one is that numerical lattice simulations are performed at finite spatial volume, so that the spectrum gets modified with respect to the infinite-volume case and the energy levels are discrete. As it will be shown below (see Eq.~(\ref{smHC2})), the $\varepsilon$-dependent kernel in Eq.\,\eqref{master0} provides a smearing of the hadronic amplitude and a necessary requirement is that, for finite values of $\varepsilon$, the smearing size is larger than the typical separation of the the discrete levels, which corresponds to the condition $\varepsilon \gg 1/L$. When this condition is satisfied, finite volume effects, which are known to be exponentially small in $\varepsilon L$~\cite{Bulava:2021fre}, can be safely kept under control.

The second observation is that, in the application of the HLT method to evaluate the smeared hadronic amplitude, the smallest values of $\varepsilon$ that can be reached depends on both the finite numbers of discrete lattice points at which the relevant time-dependent correlation function has been computed, i.e. on the temporal extension $T$ of the lattice, and on the statistical accuracy of the lattice data. Clearly,
in order to ensure good control of the extrapolation $\varepsilon \to 0$, at any selected value of the energy $E$, the condition $\varepsilon \ll \Delta(E)$ must be verified, where $\Delta(E)$ indicates the typical size of the interval around $E$ in which the hadronic amplitude is significantly varying. It may then happen that, in a region where the physical amplitude is rapidly varying, the condition $\varepsilon \ll \Delta(E)$ is difficult to be fulfilled, thus hindering in this region the extrapolation for $\varepsilon \to 0$ of the lattice results. This happens for instance when a resonance of narrow width $\Gamma$ contributes to the spectral density. In this case, in the region of the resonance $\Delta(E) \simeq \Gamma$, and the  condition $\varepsilon \ll \Gamma$ may require extremely high statistical accuracy of the lattice data. Even in this case, however, the lattice determination of the smeared hadronic amplitude at $\varepsilon>0$ may still provide valuable information, as discussed in sect.\,\ref{sec:smearedH}. For instance, if experimental data on the hadronic amplitude are available, in terms for example of a model function and experimentally determined parameters, it is then possible to evaluate from them the smeared hadronic amplitude  at $\varepsilon>0$, so that a useful comparison between theoretical and experimental data is still possible.

The method presented in this paper is general, and can be applied to the determination on the lattice of any hadronic amplitude involving two external currents and a single hadron or the QCD vacuum in the initial and final states. 
An example of process in which this problem is encountered is the radiative leptonic decays $P \to\ell\nu_\ell\, \gamma^*$, where $P$ is a pseudoscalar meson, $\ell$ is a charged lepton and $\gamma^*$ is a virtual photon produced off-shell. This process has been recently studied on the lattice in Ref.\,\cite{Gagliardi:2022szw} (see also \cite{Tuo:2021ewr}), for the case in which $P$ is a $K$ meson and the virtual photon decays into a pair of charged leptons, $\ell^{\prime\,+} \ell^{\prime\,-}$. The hadronic amplitude has been computed using the standard approach, in terms of a time-dependent Euclidean correlation function integrated over the space-time position at which the virtual photon is emitted. It is found, however, that when the virtuality $k^2$ of the photon is such that $k^2 > 4 m_\pi^2$, where $m_\pi$ is the pion mass, two-pion intermediate states with energies smaller than those of the external states contribute to the amplitude, and the integral over the Euclidean time diverges in the infinite $T$ limit. In the study of Ref.\,\cite{Gagliardi:2022szw}, addressing this problem has been postponed, by performing the numerical simulation at values of the pion mass larger than the physical value, so that the condition $k^2 > 4 m_\pi^2$ is never verified. In this work we consider the application of the spectral method to the $D_{s}\to \ell \nu_{\ell} \gamma^{*}$ decay where, neglecting the quark-line disconnected contribution and in the infinite-volume limit, the problem of analytic continuation is present for virtuality $k^{2}$ above the two-kaon threshold $4M_{K}^{2}$. For this proof-of-principle calculation we limit the simulations to a single lattice spacing $a\simeq 0.08~{\rm fm}$ and consider a single volume with spatial extent $L\simeq 5~{\rm fm}$.  

The plan for the remainder of this paper is as follows. In Sec.\,\ref{sec:method} we will discuss the method in more details, by deriving the result expressed by Eq.\,\eqref{master0} and illustrating the theoretical procedure. The numerical application of the method to the study of the $D_s \to \ell \nu_\ell \, \gamma^*$ decays will be presented in Sec.\,\ref{sec:numerics}. Finally, we end the paper by briefly presenting our conclusions and future perspectives. 

\section{The method}
\label{sec:method}
In this section we illustrate the method we are proposing in more details, deriving the general formalism and discussing some possible variants in its implementation. 

For concreteness, we will refer to the case of the hadronic amplitude of the radiative leptonic decays $P \to\ell\nu_\ell\,\gamma^*$, whose structure-dependent contribution is expressed, at the lowest-order in the electroweak interactions, by the following hadronic tensor \cite{Gagliardi:2022szw}:
\be
\label{tensor}
H_W^{\mu\nu}(k)=i\, \int d^4x\, e^{ik\cdot x}\mel{0}{T[J_{\mathrm{em}}^\mu(x) J_W^\nu(0)]}{P} \,,
\ee
where $k=(E, \bs{k})$ is the 4-momentum of the virtual photon and the initial pseudoscalar meson $P$ is taken at rest\footnote{With respect to Ref.\,\cite{Gagliardi:2022szw}, we are adopting here a different phase convention for the hadronic amplitude, defined by the $i$ in front of Eq.\,\eqref{tensor}.}. The operators $J^\mu_{\mathrm{em}}$ and $J_W^\nu$ are the electromagnetic and weak hadronic currents respectively.

The hadronic tensor $H_W^{\mu\nu}(k)$ can be split in terms of the contributions coming from the two different times-orderings, $t<0$ and $t>0$, where $x=(t,\bs{x})$ is the space-time position at which the electromagnetic current is inserted:
\bea
\label{tensor12}
H_W^{\mu\nu}(k) &=& H^{\mu\nu}_{W,1}(k)+H^{\mu\nu}_{W,2}(k) = \nn \\[0.2cm]
& = & i \, \int_{-\infty}^{0} dt \int d^3x \ e^{i E t - i \bs{k} \cdot \bs{x}}\, \mel{0}{J_W^\nu(0) J_{\mathrm{em}}^\mu(x)}{P}  \ + \\
& + & i \, \int_{0}^{\infty} dt \int d^3x \ e^{i E t - i \bs{k} \cdot \bs{x}}\, \mel{0}{J_{\mathrm{em}}^\mu(x) J_W^\nu(0)}{P} \ .  \nn
\eea
As discussed in Ref.\,\cite{Gagliardi:2022szw}, the analytic continuation from Minkowskian to Euclidean time presents no problem in the case of the first time-ordering, expressed by the hadronic tensor $H^{\mu\nu}_{W,1}(k)$. Therefore, in the following, we will concentrate the discussion on the hadronic tensor $H^{\mu\nu}_{W,2}(k)$, even though the evaluation of the hadronic tensor $H^{\mu\nu}_{W,1}(k)$ would proceed, with the spectral method we are proposing in this paper, in the same way.

In Eq.\,\eqref{tensor12}, the hadronic tensor $H^{\mu\nu}_{W,2}(k)$ is expressed in terms of the time-dependent correlator
\be
\label{eq:def_Cmunu_2}
C^{\mu\nu}(t, \bs{k}) = \int d^3x \ e^{- i \bs{k} \cdot \bs{x}}\, \mel{0}{J_{\mathrm{em}}^\mu(x) J_W^\nu(0)}{P} \qquad \quad (t>0) \ ,
\ee
so that
\be
\label{tensor2}
H_{W,2}^{\mu\nu}(k) = i\, \int_{0}^{\infty} dt \, e^{i E t} \, C^{\mu\nu}(t, \bs{k}) \ .
\ee
Here we are interested in the spectral representation of the correlator $C^{\mu\nu}(t, \bs{k})$, which can be easily derived:
 \bea
\label{Ct1}
C^{\mu\nu}(t, \bs{k}) &=& \int_{-\infty}^{+\infty} dt' \, \delta(t'-t) \, C^{\mu\nu}(t', \bs{k}) = \int_{-\infty}^{+\infty} dt' \int_{-\infty}^{+\infty} \frac{dE'}{2\pi}\, e^{i E' (t'-t)} \, C^{\mu\nu}(t', \bs{k}) = \nn \\
&=& \int_{-\infty}^{+\infty} \frac{dE'}{2\pi}\, e^{-i E' t} \int_{-\infty}^{+\infty} dt' \int d^3x' \, e^{i E' t' - i \bs{k} \cdot \bs{x'} } \mel{0}{J_{\mathrm{em}}^\mu(x') J_W^\nu(0)}{P} = \nn \\
&=& \int_{-\infty}^{+\infty} \frac{dE'}{2\pi}\, e^{-i E' t} \int d^4x' \, e^{i k' \cdot x'} \mel{0}{e^{i {\mathbb P} \cdot x'} J_{\mathrm{em}}^\mu(0) e^{-i {\mathbb P} \cdot x'} J_W^\nu(0)}{P}  = \nn \\
&=& \int_{-\infty}^{+\infty} \frac{dE'}{2\pi}\, e^{-i E' t} \int d^4x' \mel{0}{J_{\mathrm{em}}^\mu(0) e^{-i ({\mathbb P}-k') \cdot x'} J_W^\nu(0)}{P}  = \nn \\
&=& \int_{-\infty}^{+\infty} \frac{dE'}{2\pi}\, e^{-i E' t} \mel{0}{J_{\mathrm{em}}^\mu(0) \, (2\pi)^4 \delta^4({\mathbb P}-k')\,  J_W^\nu(0)}{P} \, ,
\eea
where ${\mathbb P}$ represents the 4-momentum operator and $k'=(E', \boldsymbol{k})$. Therefore, by defining the spectral density
\be
\label{spectralf}
\rho^{\mu\nu}(k) = \mel{0}{J_{\mathrm{em}}^\mu(0)\, (2\pi)^4 \delta^4({\mathbb P}-k)\, J_W^\nu(0)}{P} \, ,
\ee
and replacing the lower limit of integration on the r.h.s. of Eq.\,\eqref{Ct1} with $E^*$, where $\rho^{\mu\nu}(E', \bs{k}) =0$ for $E'<E^*$, we arrive at
\be
\label{Ctt}
C^{\mu\nu}(t, \bs{k}) = \int_{E^*}^{\infty} \frac{dE'}{2\pi}\, e^{- i E' t}\, \rho^{\mu\nu}(E', \bs{k}) \ .
\ee

Eqs.\,\eqref{spectralf} and (\ref{Ctt}) provide the spectral representation of the correlator $C^{\mu\nu}(t, \bs{k})$. While we derived these formulae for the specific case of the radiative leptonic decays of mesons, the result is general, and Eqs.\,\eqref{spectralf} and (\ref{Ctt}) hold in the same form for any process involving two external currents. In the remainder of this section, in order to keep the notation more general, we will indicate the spectral density simply as $\rho(E)$, i.e. by omitting both the Lorentz indices $\mu,\nu$ and the dependence on the spatial momentum $\bs{k}$. Similarly, we will indicate the correlation function $C^{\mu\nu}(t, \bs{k})$ with $C(t)$ and the hadronic tensor $H_{W,2}^{\mu\nu}(E, \bs{k}) $ with $H(E)$, so that Eq.\,\eqref{Ctt} is rewritten as 
\be
\label{Ct}
C(t) = \int_{E^*}^{\infty} \frac{dE'}{2\pi}\, e^{- i E' t}\, \rho(E') \ .
\ee

The spectral representation of the hadronic tensor $H(E)$ is now derived using Eqs.\,\eqref{tensor2} 
and \eqref{Ct} to obtain
\bea
\label{regulator}
H(E) &=& i\, \int_{0}^{\infty} dt \, e^{i E t} \, C(t) \, = \, i\, \int_{0}^{\infty} dt \int_{E^*}^{\infty} \frac{dE'}{2\pi}\, e^{-i (E' -E) t} \, \rho(E') = \nn \\ 
&=& i \,\lim_{\varepsilon \to 0} \, \int_{E^*}^{\infty} \frac{dE'}{2\pi}  \, \rho(E') \int_{0}^{\infty} dt \, e^{-i (E' -E - i \varepsilon) t} \ ,
\eea
where, in the last equality, a factor $e^{-\varepsilon t}$, with $\varepsilon > 0$ and $\varepsilon \to 0$, has been introduced in the time integral in order to ensure the convergence of the integral for $t \to \infty$. Finally, by performing the integral over time in Eq.\,\eqref{regulator}, one arrives to the expression
\be
\label{master}
H(E) = \lim_{\varepsilon \to 0} \, \int_{E^*}^{\infty} \frac{dE'}{2\pi} \, \frac{\rho(E') }{E' -E - i \varepsilon} \ ,
\ee
which relates the hadronic amplitude $H(E)$ to the spectral density $\rho(E')$. Eq.\,\eqref{master}, which has been anticipated in Eq.\,\eqref{master0}, represents the basic ingredient of the method we are proposing in this paper to evaluate the hadronic amplitude on the lattice. 

It is important to note that Eq.\,\eqref{master} does not depend on time. Therefore, it can be used to evaluate the hadronic amplitude through a lattice calculation performed on an Euclidean space-time, without encountering any difficulty related to the analytical continuation from real to imaginary time. The only (trivial) analytic continuation which has to be considered concerns the relation in Eq.\,\eqref{Ct} between the correlator and its spectral density. Once expressed in terms of the Euclidean time, this relation takes the form
\be
\label{CtE}
C_E(t) = \int_{E^*}^{\infty} \frac{dE'}{2\pi}\, e^{- E' t}\, \rho(E') \ ,
\ee
where $C_E(t)$ denotes the Euclidean correlator.

The structure of the hadronic amplitude $H(E)$ which follows from Eq.\,\eqref{master} is simpler in the case in which all the internal states contributing to the spectral function $\rho(E)$ have energies $E_n > E$ (so that also $E^* > E$). In this case, the denominator in the integrand of Eq.\,\eqref{master} has no poles and the limit $\varepsilon \to 0$ can be taken directly, leading to
\be
H(E) = \int_{E^*}^{\infty} \frac{dE'}{2\pi} \, \frac{\rho(E') }{E' -E} \ , \qquad \textrm{for}~ E^* > E \ .
\ee
Then, by replacing
\be
\frac{1}{E' -E} = \int_{0}^{\infty} dt \, e^{-(E' -E) t} \ ,
\ee
which is valid for $E' > E$, and using Eq.\,\eqref{CtE}, one arrives at
\bea
\label{ampE}
H(E) &=& \int_{0}^{\infty} dt \int_{E^*}^{\infty} \frac{dE'}{2\pi} \, e^{-(E' -E) t} \, \rho(E') = \nn \\
 &=& \int_{0}^{\infty} dt \, e^{E t} \, C_E(t)  \ ,
\eea
which expresses the hadronic amplitude in terms of the Euclidean correlator when all the states contributing to the spectral function $\rho(E)$ have energies $E_n > E$. Eq.\,\eqref{ampE} has been used in Ref.\,\cite{Gagliardi:2022szw} to evaluate the hadronic amplitude on the lattice, and represents what we refer to in this paper as the standard approach.

When however one or more states contributing to the spectral function $\rho(E)$ have energies $E_n < E$, then the denominator in the integrand of Eq.\,\eqref{master} develops singularities for $\varepsilon = 0$. This implies that the unregularized Euclidean integral of Eq.\,\eqref{ampE} is divergent and the Wick rotation from Minkowskian to Euclidean time could have not been performed. In this case, however, the regularized spectral representation of Eq.\,\eqref{master} is still valid, provided that the limit $\varepsilon \to 0$ is taken only after the integral has been evaluated. Therefore, for finite values of $\varepsilon$, the integral of Eq.\,\eqref{master} can be used to evaluate the hadronic amplitude in a lattice simulation.

When intermediate states with energies smaller than $E$ contribute to the spectral function, the hadronic amplitude $H(E)$ becomes complex, due to the presence of the $i \varepsilon$ term in the denominator of Eq.\,\eqref{master}. Explicitly, its real and imaginary parts are expressed by\footnote{ In writing down Eqs.~(\ref{masterR})-(\ref{masterI}), and in the following, we assume that the spectral density $\rho(E')$ is real. If this is not the case, the formulae describing the separation of the hadronic amplitude in its real and imaginary part admit a straightforward generalization. }
\bea
&& \Re H(E) \, = \, 
\lim_{\varepsilon \to 0} \, \int_{E^*}^{\infty} \frac{dE'}{2\pi} \left[ \frac{(E'-E)}{(E'-E)^2 + \varepsilon^2} \right] \, \rho(E') \ ,
\label{masterR}
 \\[0.2cm]
&& \Im H(E) \, = \, 
\lim_{\varepsilon \to 0} \, \int_{E^*}^{\infty} \frac{dE'}{2\pi} \left[ \frac{\varepsilon}{(E'-E)^2 + \varepsilon^2} \right] \, \rho(E') 
\ .
\label{masterI}
\eea
Our proposal, as in Ref.~\cite{Bulava:2019kbi}, consists in determining the real and the imaginary part of the hadronic amplitude by computing the integrals in Eqs.\,\eqref{masterR} and \eqref{masterI}, at finite values of $\varepsilon$, using the HLT method of Ref.\,\cite{Hansen:2019idp} (see also Ref.\,\cite{Alexandrou:2022tyn}), starting from the results for the Euclidean correlator $C_E(t)$ evaluated non-perturbatively on the lattice. The main ingredients of the HLT method will be summarized in sect.\,\ref{sec:numerics}, together with the results of an exploratory numerical calculation. Before coming to that, however, we further elaborate on the theoretical proposal.

\subsection{Alternative choices of the kernel}
\label{sec:kernels}

In the limit $\varepsilon \to 0$, the kernel functions in Eqs.\,\eqref{masterR} and \eqref{masterI} lead respectively to the principal value of the integral and to a delta function, so that
\bea
&& \Re H(E) \, = \, PV \int_{E^*}^{\infty} \frac{dE'}{2\pi} \ \frac{1}{(E' -E)} \ \rho(E') \ ,
\label{masterR0} 
\\[0.2cm]
&& \Im H(E) \, = \, \int_{E^*}^{\infty} \frac{dE'}{2\pi} \ \pi\, \delta(E' -E) \ \rho(E') \, = \, \frac{1}{2} \, \rho(E) \ ,
\label{masterI0} 
\eea
where $PV$ denotes the principal value. It should be then clear that the spectral representation \eqref{master} of the hadronic amplitude can be expressed in the more general form 
\be
\label{master1}
H(E) = \lim_{\varepsilon \to 0} \, \int_{E^*}^{\infty} \frac{dE'}{2\pi} \, K(E'-E;\, \varepsilon) \, \rho(E') \ ,
\ee
where the kernel $K(E;\, \varepsilon)$ is any complex function that, in the limit $\varepsilon \to 0$, reproduces Eqs.\,\eqref{masterR0} and \eqref{masterI0}. 

The choice made in Eq.\,\eqref{master}, corresponds to the Cauchy kernel
\be
\label{CauchyK}
K^C(E;\, \varepsilon) = \frac{1}{E-i\varepsilon} \ .
\ee

An alternative choice of the kernel is obtained by regularizing the time integral in Eq.\,\eqref{regulator}  with a gaussian factor $\exp(- \varepsilon^{2}\, t^2/2)$ instead of $\exp(- \varepsilon\, t)$. In this way, one obtains for the hadronic amplitude the expression
\be
\label{gaussreg1}
H(E) = i\, \lim_{\varepsilon \to 0} \, \int_{E^*}^{\infty} \frac{dE'}{2\pi}  \, \rho(E') \int_{0}^{\infty} dt \, e^{-i (E' -E)t - \varepsilon^{2} \,t^2/2} \ .
\ee
The integral over time can be performed analitically and after a simple algebra one gets, for the real and imaginary parts of the kernel, the expressions
\be
\label{kernelsRIg}
\Re K^g(E; \varepsilon) \, = \, \frac{\sqrt 2}{\varepsilon}\, D_+\left( \frac{E}{\sqrt2 \, \varepsilon} \right) \quad , \quad
\Im K^g(E; \varepsilon) \, = \, \sqrt{\frac{\pi}{2\, \varepsilon^2}} \, \exp\left[- \left( \frac{E}{\sqrt2 \, \varepsilon} \right)^2\right]  \ ,
\ee
where 
\be
D_+(x) = e^{-x^2} \, \int_{0}^{x} ds \, e^{s^2} 
\ee
is the so called Dawson function.

In the numerical lattice calculation presented in sect.\,\ref{sec:numerics} we find convenient to adopt for the kernel a lattice discretized version of the Cauchy kernel of Eq.\,\eqref{CauchyK}. This kernel depends on the lattice spacing $a$ and, in the limit $\varepsilon \to 0$, it leads for the hadronic amplitude to the same result obtained on the lattice at finite lattice spacing with the standard approach (in the region $E < E^*$ where the standard approach is applicable). In order to derive the expression of this kernel, we start from the lattice discretized version of Eq.\,\eqref{ampE}, in which the Euclidean time is restricted to assume multiple integer values of the lattice spacing, $t=a\,n$ with $n=1,2,\ldots , T/(2a)$, and the integral over time is thus replaced by a finite sum. Using
\be
\sum_{n=1}^{N}\, x^n = \frac{x\, (1-x^N)}{1-x} \ ,
\ee
we find:
\bea
\label{ampElat}
H(E) &=& a\, \sum_{n=1}^{T/(2a)} \ e^{a E n} \ C_E(a n) = a\, \sum_{n=1}^{T/(2a)} \ \int_{E^*}^{\infty} \frac{dE'}{2\pi} \,  e^{-a (E' -E) n} \, \rho(E')  = \nn \\
&=& a\, \int_{E^*}^{\infty} \frac{dE'}{2\pi} \,  \frac{e^{-a (E' -E)}}{1-e^{-a (E' -E)}} \left(
1- e^{-(E' -E)\, T/2} \right) \rho(E') = \nn \\
&=& a\, \int_{E^*}^{\infty} \frac{dE'}{2\pi} \,  \frac{e^{-a (E' -E)/2}}{2\sinh{[a (E' -E)/2]}} \left(1- e^{-(E' -E)\, T/2} \right) \rho(E')
\ .
\eea
It can be seen again, from the previous expression, that, if there are intermediate states contributing to the spectral function $\rho$ which have energies smaller than the energy $E$ of the external states, i.e. $E^*<E$, then the integral in Eq.\,\eqref{ampElat} receives the contribution of terms which are
exponentially large in $T$, and diverge in the $T \to \infty$ limit. For $E<E^*$, instead, these terms are exponentially small, and can be safely neglected for large $T$, leading for the hadronic amplitude to the expression
\be
\label{ampElat2}
H(E) = a\, \int_{E^*}^{\infty} \frac{dE'}{2\pi} \,  \frac{e^{-a (E' -E)/2}}{2\sinh{[a (E' -E)/2]}} \ \rho(E')
\quad , \quad \textrm{for}~E<E^* \ .
\ee

The comparison between Eqs.\,\eqref{ampElat2} and \eqref{master1} shows that the lattice evaluation of the hadronic amplitude with the standard approach for $E<E^*$, i.e. in terms of the time dependent correlation function, is equivalent, at finite lattice spacing, to use the spectral method with the kernel 
\be
\label{Cauchylat0}
K^L(E;\, \varepsilon=0) = 
\frac{ e^{-a E/2}}{(2/a) \sinh{[a E/2]}} \ .
\ee
It is then natural to define a lattice version of the kernel at $\varepsilon \neq 0$ as
\be
\label{Cauchylat}
K^L(E;\, \varepsilon) = K^{L}(E-i\varepsilon; 0) =  
\frac{e^{-a (E-i\varepsilon)/2}}{(2/a) \sinh{[a (E-i\varepsilon)/2]}} \ ,
\ee
which reduces to the Cauchy kernel \eqref{CauchyK} in the continuum limit\footnote{We notice that the lattice Cauchy kernel in Eq.~(\ref{Cauchylat}) differs from the continuum kernel by a term of $\mathcal{O}(a)$. This introduces an $\mathcal{O}(a)$ correction in the amplitude~(\ref{ampElat}), which represents a lattice discretization of $H^{\mu\nu}_{W,2}$. Such an $\mathcal{O}(a)$ artifact would cancel upon combining this lattice discretization of $H^{\mu\nu}_{W,2}$ with the appropriate lattice discretization of $H^{\mu\nu}_{W,1}$. No $\mathcal{O}(a)$ artifacts are thus expected after properly combining the two time orderings contributing to the physical amplitude in Eq.~(\ref{tensor12}).}$a \to 0$. As shown by the above discussion, with this choice of the kernel the lattice artefacts affecting the hadronic amplitude obtained with either the standard or the spectral approach at finite lattice spacing, for  $E<E^*$, are exactly equal (up to exponentially small corrections vanishing for $T \to \infty$). This feature 
turns out to be beneficial for the comparison between the results obtained, at fixed lattice spacing, with the new and the standard method, in the region $E<E^*$ where the latter is applicable.

\subsection{The smeared amplitude}
\label{sec:smearedH}

In the region $E>E^*$, where the spectral representation of the hadronic amplitude requires a regularization, the integral over energy of  Eq.\,\eqref{master1} can be only evaluated at finite values of the smearing parameter $\varepsilon$, and it is only at the end of the calculation that the extrapolation of the results at $\varepsilon \to 0$ can be eventually performed. Therefore, the quantity which is directly obtained from the lattice calculation is the {\em smeared amplitude} at finite $\varepsilon$,
\be
\label{smH}
H_K(E;\varepsilon) = \int_{E^*}^{\infty} \frac{dE'}{2\pi} \, K(E'-E;\, \varepsilon) \, \rho(E') \ .
\ee
According to Eq.\,\eqref{master1}, and independently of the specific choice of the kernel, the smeared amplitude tends to the physical hadronic amplitude in the limit $\varepsilon \to 0$,
\be
\label{smH2}
H(E) = \lim_{\varepsilon \to 0} \,  H_K(E;\varepsilon) \ .
\ee

It is interesting to note that, in the case of the Cauchy kernel \eqref{CauchyK}, the smeared amplitude
\be
\label{smHC1}
 H_C(E;\varepsilon) = \int_{E^*}^{\infty} \frac{dE'}{2\pi} \, \frac{\rho(E')}{E'-E-i \varepsilon}
\ee
can be directly expressed in terms of the physical amplitude $H(E)$, i.e.
\be
\label{smHC2}
 H_C(E;\varepsilon) = \int_{-\infty}^{+\infty} \frac{dE'}{2\pi} \, \frac{2\, \varepsilon}{(E'-E)^2 + \varepsilon^2}\, H(E') \ ,
\ee
as it can be proven by substituting Eqs.\,\eqref{smH2} and \eqref{smHC1} into Eq.\,\eqref{smHC2} and using
\be
\lim_{\eta \to 0}\, \frac{1}{\pi}\,\int_{-\infty}^{+\infty} d\omega \ \frac{\varepsilon}{(E-\omega)^2 + \varepsilon^2}\ \frac{1}{E'-\omega-i \eta} \, = \, \frac{1}{E'-E-i \varepsilon} \ .  
\ee
Eq.\,\eqref{smHC2} clearly shows that the smeared amplitude $ H_C(E;\varepsilon)$ represents in fact a smearing of size $\varepsilon$ of the physical hadronic amplitude $H(E)$. In addition, by providing a direct relation between the smeared amplitude and the physical one, independent of the knowledge of the spectral density $\rho$, Eq.\,\eqref{smHC2} may be also useful in the phenomenological analyses of lattice results. Indeed, by having a model expression for the physical amplitude $H(E)$, based on theoretical or phenomenological considerations, Eq.\,\eqref{smHC2} allows to evaluate the model-dependent smeared hadronic amplitude $H_C(E;\varepsilon)$  and to compare it directly with the results of the lattice calculation at finite $\varepsilon$. An example of the utility of such analysis in the proximity of a narrow resonance is discussed in sec.\,\ref{sec:eps_extrapolation}.

For illustration, let us consider a simple model of a resonance of mass $M$ and width $\Gamma$. In this case, the spectral density reads
\be
\label{eq:mod}
\rho(E) = \frac{2A\,\Gamma}{(E-M)^2 + \Gamma^2}\,\theta(E) \ ,
\ee
where $A$ is a constant, and the theta function ensures that the spectral density vanishes for negative energies. For sufficiently large values of $M$ and small values of $\Gamma$, the effect of the theta function is practically negligible, and for the sake of simplicity it will be neglected. 

\begin{figure}
    \centering
\includegraphics[scale=0.4]{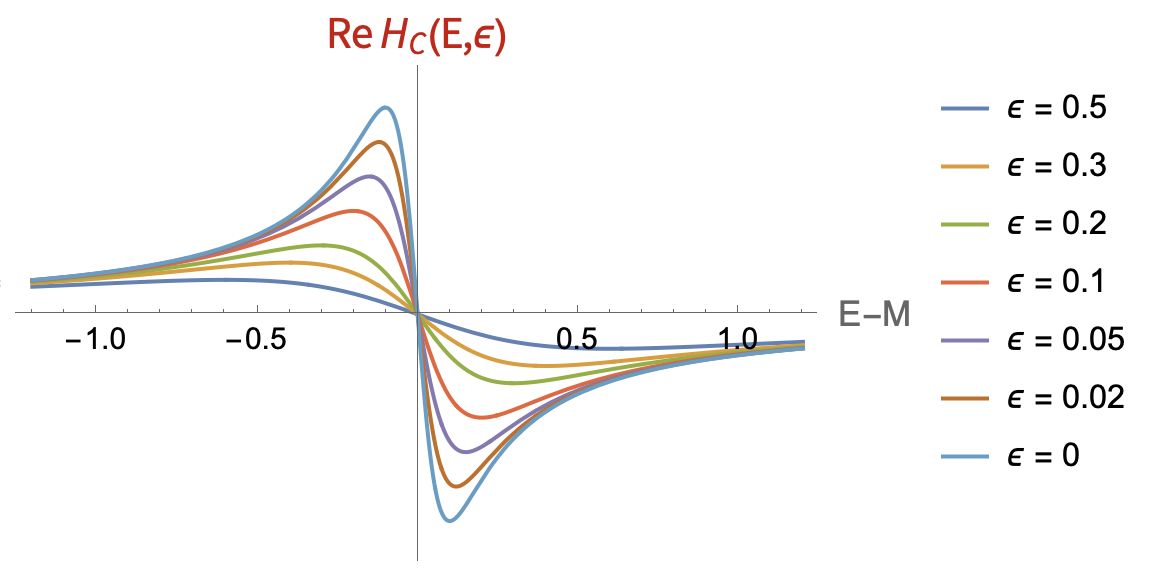} \hspace{0.5cm}
\includegraphics[scale=0.4]{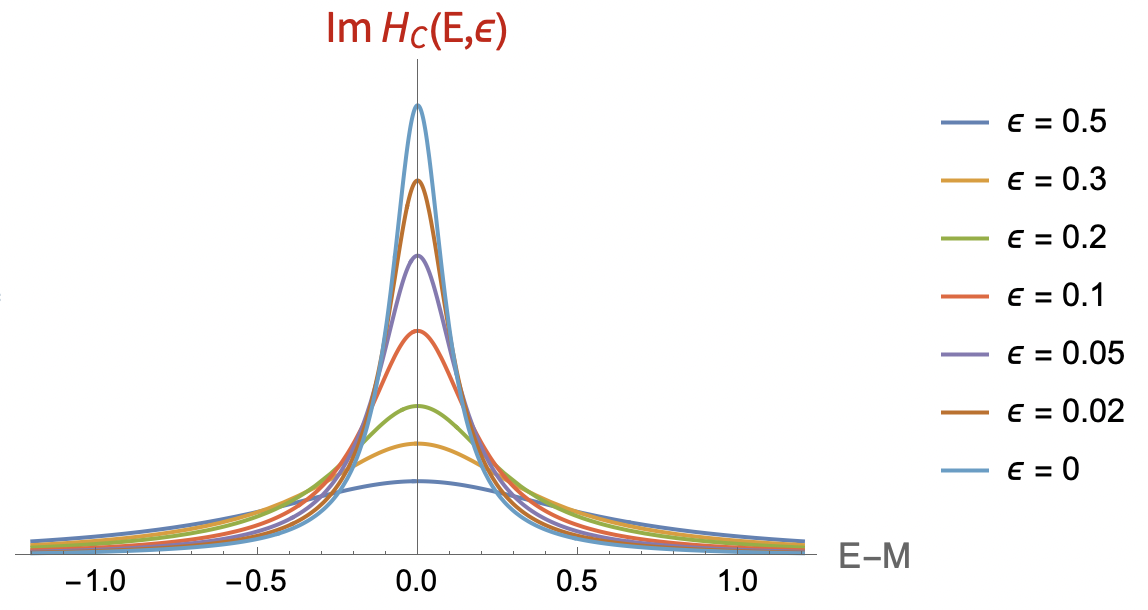}
    \caption{\it\small The real (left) and imaginary (right) part of the smeared amplitude $H_C(E;\varepsilon)$ of Eq.\,\eqref{smH-model} (in arbitrary units) for the single resonance model, for different values of $\varepsilon$ and $\Gamma=0.1$. The curve corresponding to $\varepsilon=0$ represents the physical amplitude.}
    \label{fig:smH-model}
\end{figure}

The time-dependent Minkowskian correlator for this model is obtained by integrating the spectral density $\rho(E)$ according to Eq.\,\eqref{Ct}, and one finds
\be
C(t) = A\, e^{-i\, (M - i\, \Gamma)\, t} \ .
\ee
The smeared amplitude $H_C(E;\varepsilon)$, defined with the Cauchy kernel, can be computed in terms of the spectral density using Eq.\,\eqref{smHC1}, and one finds
\be
\label{smH-model}
H_C(E;\varepsilon) = - \frac{A}{E-M + i\, (\Gamma + \varepsilon)} \ .
\ee
As we can see, the effect of the smearing in this case consists in replacing the width of the resonance $\Gamma$ with $\Gamma + \varepsilon$. The real and imaginary parts of the smeared amplitude \eqref{smH-model} are given by
\be
\label{smH-modelRI}
\Re H_C(E;\varepsilon) = - \frac{A\, (E-M)}{(E-M)^2 + (\Gamma + \varepsilon)^2} \quad , \quad 
\Im H_C(E;\varepsilon) = \frac{A\, (\Gamma + \varepsilon)}{(E-M)^2 + (\Gamma + \varepsilon)^2} 
\ee
and are shown, for illustrative purpose, in Fig.\,\ref{fig:smH-model} for different values of the smearing size, ranging from $\varepsilon=0.5$ down to $\varepsilon=0$.

In the limit $\varepsilon\to 0$, the smeared amplitude of Eq.\,\eqref{smH-model} tends to the physical amplitude
\be
\label{H-model}
H(E) = - \frac{A}{E-M + i\, \Gamma} \ .
\ee
It can be easily seen that the real and imaginary parts of the amplitude satisfy
\bea
\Re H(E) &=& - \frac{A\, (E-M)}{(E-M)^2 + \Gamma^2} =
\int_{-\infty}^{+\infty} \frac{dE'}{2\pi} \ 
\frac{\rho(E+E')-\rho(E-E')}{2\, E'} = 
PV \int_{-\infty}^{+\infty} \frac{dE'}{2\pi} \ \frac{\rho(E+E')}{E'} \ , \\[0.2cm]
\Im H(E) &=& \frac{A\, \Gamma}{(E-M)^2 + \Gamma^2} =
\frac{1}{2} \, \rho(E) \ ,
\eea
in agreement with Eqs.\,\eqref{masterR0} and \eqref{masterI0}. One can also verify that the smeared amplitude \eqref{smH-model} can be obtained directly from the physical amplitude \eqref{H-model} using Eq.\,\eqref{smHC2}.

\subsection{The \texorpdfstring{$\varepsilon$}{epsilon}-expansion}
\label{sec:epsexpansion}

As already discussed in the Introduction, the smallest values of the smearing size $\varepsilon$ that can be reached in the numerical calculation of the smeared hadronic amplitude with the HLT method depend on both the number of discrete lattice times at which the relevant correlation function $C(t)$ is known, i.e. on the temporal extension $T$ of the lattice, and on the statistical accuracy of the lattice data. In order to have good control of the final extrapolation to $\varepsilon \to 0$, it is then instructive to understand the $\varepsilon$-dependence of the smeared hadronic amplitude $ H_K(E;\varepsilon)$. We present here this analysis for the case of the Cauchy kernel \eqref{CauchyK}. Since the lattice kernel of Eq.\,\eqref{Cauchylat} reduces to the Cauchy kernel in the limit $a \to 0$, the same results will be also valid for the lattice kernel up to finite lattice artefacts vanishing for $a\to 0$.

Let us consider the smeared hadronic amplitude of Eq.\,\eqref{smHC1}. By noting that the spectral density vanishes for $E<E^*$, we can replace the lower integration limit in Eq.\,\eqref{smHC1} with $-\infty$. We can then make the change of variable $\omega = E'-E$ and obtain, for the real and imaginary parts of the smeared amplitude, the expressions 
\bea
&& \Re  H_C(E;\varepsilon) \, = \, 
 \int_{-\infty}^{+\infty} \frac{d\omega}{2\pi} \left( \frac{\omega}{\omega^2 + \varepsilon^2} \right) \, f(\omega) \ ,
\label{expanR1}
 \\[0.2cm]
&& \Im  H_C(E;\varepsilon) \, = \, 
 \int_{-\infty}^{+\infty} \frac{d\omega}{2\pi} \left( \frac{\varepsilon}{\omega^2 + \varepsilon^2} \right) \, f(\omega) \ ,
\label{expanI1}
\eea
where we have put, for brevity, $f(\omega) = \rho(E+\omega)$. 

In the real part of the smeared amplitude, the kernel is an odd function of $\omega$. Therefore, Eq.\,\eqref{expanR1} can be rewritten in the form:  
\bea
\Re  H_C(E;\varepsilon) &=& 
 \int_{-\infty}^{+\infty} \frac{d\omega}{2\pi} \left( \frac{\omega}{\omega^2 + \varepsilon^2} \right) \, \frac{1}{2} \left( f(\omega) -f(-\omega) \right) = \nn \\[0.2cm]
 &=& \int_{-\infty}^{+\infty} \frac{d\omega}{2\pi} \left( \frac{\omega^2 + \varepsilon^2 - \varepsilon^2}{\omega^2 + \varepsilon^2} \right) \, \frac{f(\omega) -f(-\omega)}{2\, \omega} =\nn \\[0.2cm]
 &=& \int_{-\infty}^{+\infty} \frac{d\omega}{2\pi} \, \frac{f(\omega) -f(-\omega)}{2\, \omega} \, - \, \int_{-\infty}^{+\infty} \frac{d\omega}{2\pi} \left( \frac{\varepsilon^2}{\omega^2 + \varepsilon^2} \right) \, \frac{f(\omega) -f(-\omega)}{2\, \omega} \ .
\label{expanR2}
\eea
The first term on the r.h.s. of Eq.\,\eqref{expanR2} is independent of $\varepsilon$ and provides, as expected, the principal value of the integral of $f(\omega)/\omega$, i.e. the real part of the physical amplitude. The second term represents instead the correction for finite values of $\varepsilon$. By reintroducing $f(\omega) = \rho(E+\omega)$, we thus obtain
\be
 \Re  H_C(E;\varepsilon) =
  PV \int_{-\infty}^{+\infty} \frac{d\omega}{2\pi} \, \frac{\rho(E+\omega)}{\omega} \, - \, \varepsilon \, \int_{-\infty}^{+\infty} \frac{d\omega}{2\pi} \left( \frac{\varepsilon}{\omega^2 + \varepsilon^2} \right) \, \frac{\rho(E+\omega) -\rho(E-\omega)}{2\, \omega} \ .
\label{expanR3}
\ee
A similar manipulation can be carried out for the imaginary part the smeared amplitude. In this case, the kernel in Eq.\,\eqref{expanI1} is an even function of $\omega$ and we can write:  
\bea
\Im  H_C(E;\varepsilon) &=& 
 \int_{-\infty}^{+\infty} \frac{d\omega}{2\pi} \left( \frac{\varepsilon}{\omega^2 + \varepsilon^2} \right) \, \frac{1}{2} \left( f(\omega) + f(-\omega) \right) = \nn \\[0.2cm]
 &=& \frac{1}{2} \, \int_{-\infty}^{+\infty} \frac{d\omega}{2\pi} \left( \frac{\varepsilon}{\omega^2 + \varepsilon^2} \right) \, \left( f(\omega) + f(-\omega) -2\, f(0) + 2\, f(0) \right) = \nn \\[0.2cm]
 &=& \frac{1}{2} \, f(0) \, + \, \frac{1}{2} \, \varepsilon \int_{-\infty}^{+\infty} \frac{d\omega}{2\pi} \left( \frac{1}{\omega^2 + \varepsilon^2} \right) \, \left(f(\omega) + f(-\omega) -2\, f(0)\right) \ ,
\label{expanI2}
\eea
or, by reinserting $f(\omega) = \rho(E+\omega)$,
 \be
\Im  H_C(E;\varepsilon) = \frac{1}{2} \, \rho(E) \, + \,  \frac{1}{2} \, \varepsilon \int_{-\infty}^{+\infty} \frac{d\omega}{2\pi} \left( \frac{1}{\omega^2 + \varepsilon^2} \right) \, \left( \rho(E+\omega) + \rho(E-\omega) -2\, \rho(E) \right) \ .
\label{expanI3}
\ee
Again, the first term on the r.h.s. of Eq.\,\eqref{expanI3} is independent of $\varepsilon$ and provides the imaginary part of the physical hadronic amplitude. The second term represents the correction for finite values of $\varepsilon$. 

We can rewrite Eqs.\,\eqref{expanR3} and \eqref{expanI3} in a more compact and explanatory form as
\bea
 \Re H_C(E;\varepsilon) &=& \Re H(E) \, - \, \varepsilon \, \int_{-\infty}^{+\infty} \frac{d\omega}{2\pi} \left( \frac{\varepsilon}{\omega^2 + \varepsilon^2} \right) \, \rho^{[1]}(E,\omega) \ , \\[0.4cm]
\label{expanR4}
\Im H_C(E;\varepsilon) &=& \Im H(E)  \, + \,  \frac{1}{2} \, \varepsilon \int_{-\infty}^{+\infty} \frac{d\omega}{2\pi} \left( \frac{\omega}{\omega^2 + \varepsilon^2} \right) \, \omega \, \rho^{[2]}(E,\omega) \ ,
\label{expanI4}
\eea
where we have defined the functions
\bea
\rho^{[1]}(E,\omega) &=& \frac{\rho(E+\omega) -\rho(E-\omega)}{2\, \omega} \ , \nn \\[0.2cm]
\rho^{[2]}(E,\omega) &=&
\frac{\rho(E+\omega) + \rho(E-\omega) -2\, \rho(E)}{\omega^2} \ .
\eea
It is useful to note that the $\varepsilon$-dependent correction term in the real part of the smeared amplitude is expressed as an integral with the same kernel of the imaginary part, and vice versa. For small values of $\varepsilon$, Eqs.\,\eqref{expanR3} and \eqref{expanI3} can thus be expanded iteratively in powers of $\varepsilon$. In particular, at the leading order in $\varepsilon$ and by assuming that the first derivative of $\rho(\omega)$ at $\omega = E$ exists, one finds
\bea
\label{eq:exp_RE}
\Re  H_C(E;\varepsilon) &=& \Re H(E) \, - \, \frac{1}{2}\, \varepsilon \, \rho'(E) \, + \, {\cal O}(\varepsilon^2) \ , \\[0.2cm]
\label{eq:exp_IM}
\Im  H_C(E;\varepsilon) &=& \Im H(E) \, + \, \frac{1}{2} \, \varepsilon \, \int_{-\infty}^{+\infty} \frac{d\omega}{2\pi} \ \rho^{[2]}(E,\omega) \, + \, {\cal O}(\varepsilon^2) \ .
\eea
An immediate consequences of Eq.\,\eqref{eq:exp_RE} is that the ${\cal O}(\varepsilon)$-correction to the real part of the smeared amplitude vanishes in the region $E<E^*$, where the spectral density $\rho(E)$, and therefore also its derivative $\rho'(E)$, vanish.

It is instructive to look at the $\varepsilon$-expansion of the smeared amplitude in the case of the one-resonance model discussed in the previous subsection, for which the smeared amplitude $H_C(E, \varepsilon)$ is given by Eq.\,\eqref{smH-model}. It can be seen that the ratio between $H_C(E, \varepsilon)$ and the physical amplitude $H(E)$ in this model can be written in the form
\be
\label{smHexp}
\frac{H_C(E, \varepsilon)}{H(E)} = \left[ 1 + i\, \varepsilon \, \frac{(E-M) - i \, \Gamma}{(E-M)^2 + \Gamma^2} \right]^{-1} =  \left[ 1 + \frac{i\, \varepsilon}{\Delta (E)} \, e^{-i \phi(E)} \right]^{-1} \ ,
\ee
where 
\be
\label{DeltaePhi}
\Delta(E) = \sqrt{(E-M)^2 + \Gamma^2} \quad , \quad 
\tan \phi(E) = \frac{\Gamma}{E-M} \ .
\ee
It then clear from Eq.\,\eqref{smHexp} that the $\varepsilon$-expansion of $H_C(E,\varepsilon)$ is effectively an expansion in powers of the dimensionless parameter $\varepsilon/\Delta(E)$, and a fast convergence of the expansion is thus expected for $\varepsilon \ll \Delta(E)$. Note that the energy $\Delta(E)$ is related to the logarithmic derivative of the hadronic amplitude by
\be
\label{deltaEdef}
 \left| \frac{\partial\log( H(E))}{\partial E} \right| =  \left| \frac{1}{H(E)}\frac{ \partial H(E)}{\partial E} \right| = \frac{1}{\Delta(E)} \ ,
\ee
implying that $\Delta(E)$ is the size of the interval in which the amplitude $H(E)$ varies by $\mathcal{O}(100\%)$,
\be
 \left| \frac{ \partial H(E)}{\partial E} \right| \Delta(E) = \left| H(E) \right| \ .
\ee
Therefore, the condition $\varepsilon \ll \Delta(E)$, which we had anticipated in the Introduction as the 
requirement for a good convergence of the $\varepsilon$-expansion, can be read in general, beyond the specific model we are now discussing, by formally defining $\Delta(E)$ from Eq.\,\eqref{deltaEdef}.

Looking at the expansion of the real and imaginary parts of the smeared amplitude in the one-resonance model, we find, at the leading order,  
\bea
\Re H_C(E, \varepsilon) &=& - \frac{A\, (E-M)}{(E-M)^2 + (\Gamma + \varepsilon)^2} = 
\Re H(E) + \frac{2 A\, (E-M)\,\Gamma}{[(E-M)^2 + \Gamma^2]^2}\, \varepsilon + {\cal O}(\varepsilon^2) \\[0.2cm]
\Im H_C(E, \varepsilon) &=& \frac{A\, (\Gamma + \varepsilon)}{(E-M)^2 + (\Gamma + \varepsilon)^2} = 
\Im H(E) + \frac{A\, [(E-M)^2 - \Gamma^2]}{[(E-M)^2 + \Gamma^2]^2}\, \varepsilon + {\cal O}(\varepsilon^2)
\eea
which can be seen to be in agreement with the general result of Eqs.\,\eqref{eq:exp_RE} and \eqref{eq:exp_IM}. In terms of $\Delta(E)$ and $\phi(E)$ of Eq.\,\eqref{DeltaePhi}, the previous expression can be written in the simple form
\bea
\Re H_C(E, \varepsilon) &=&  \Re H(E) + A\, \frac{\sin{(2\, \phi(E))} }{\Delta(E)^2}\, \varepsilon + {\cal O}(\varepsilon^2) \\[0.2cm]
\Im H_C(E, \varepsilon) &=& 
\Im H(E) + A\, \frac{\cos{(2\, \phi(E))} }{\Delta(E)^2} \, \varepsilon + {\cal O}(\varepsilon^2) \ ,
\eea
which show again that the $\varepsilon$-expansion is effectively an expansion in powers of the dimensionless ratio $\varepsilon/\Delta(E)$.

\section{The numerical application}
\label{sec:numerics}
In this section we present the application of the spectral method to a realistic numerical calculation. We consider the radiative leptonic decay $D_s \to \ell \nu \gamma^*$ and aim to calculate the hadronic tensor $H^{\mu\nu}_{W,2}$ defined in Eqs.~(\ref{eq:def_Cmunu_2})-~(\ref{tensor2}), with $P=D_{s}$. We perform the calculation for several value of the energy of the virtual photon, which cover both the region $E<E^*$, where the standard approach based on Eq.\,\eqref{ampE} is also applicable, and the region $E>E^*$ where instead the Euclidean integral of Eq.\,\eqref{ampE} becomes divergent for $T \to \infty$ and we rely on the new method based on the spectral representation.

\begin{figure}
    \centering
    \includegraphics[scale=0.8]{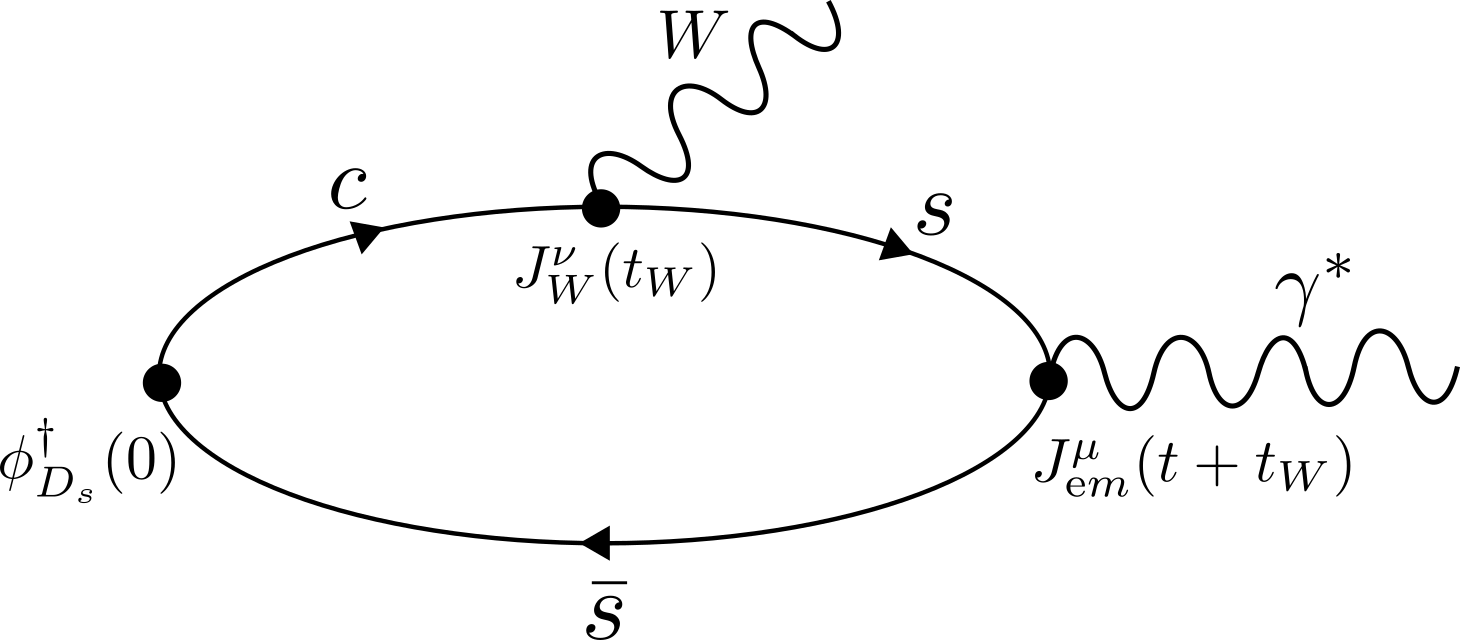}
    \caption{\it\small Graphical representation of the three-point Euclidean function in Eq.~(\ref{eq:M_munu}). The interpolator of the $D_{s}$ meson $\phi^{\dag}_{D_{s}}$ is placed at Euclidean time $0$, the weak current $J_{W}^{\nu}$ at time $t_{W}$ and the electromagnetic current at time $t+t_{W}$. In the figure we show only the contribution where the virtual photon $\gamma^{*}$ is emitted from the strange quark $s$, but the analogous diagram with the photon emitted from the charm quark $c$ is clearly present as well.}
    \label{fig:Feyn}
\end{figure}

For the present study, we make use of a single ensemble generated by the Extended Twisted Mass Collaboration (ETMC) employing the Iwasaki gluon action and $N_{f}=2+1+1$ flavours of Wilson-Clover twisted-mass fermions at maximal twist~\cite{Frezzotti:2000nk}. A detailed description of the ETMC ensembles can be found in Refs.~\cite{ExtendedTwistedMass:2021gbo,ExtendedTwistedMass:2021qui}, while essential information on the ensemble we have used in the present work is collected in Table\,\ref{tab:simudetails}. 
\begin{table}
    \begin{tabular}{||c||c|c|c|c|c|c|c| c||}
    \hline
    ~~~ ensemble ~~~ & ~~~ $\beta$ ~~~ & ~~~ $V/a^{4}$ ~~~ & ~~~ $a$\,(fm) ~~~ & ~ $M_{\pi}$\,(MeV) ~ & ~ $M_{D_{s}}$(GeV) ~ & ~ $L$ (fm)  ~ & ~ $N_{g}$ ~  & ~ $N_{\rm s}$ ~\\
  \hline  
  cB211.072.64 & $1.778$ & $64^{3}\cdot 128$ & $0.07957~(13)$ & $140.2~(0.2)$ & $1.990~(3)$  & $5.09$ & $300$ & 4  \\
  \hline
    \end{tabular}
\caption{\it \small Parameters of the single ETMC ensemble used in this work. We give the lattice spacing $a$,  the pion mass $M_\pi$, the $D_{s}$ meson mass $M_{D_{s}}$, the lattice extent $L$, the number of gauge configurations analyzed $N_{g}$, and the number $N_{\rm s}$ of random stochastic sources that have been used for each inversion of the Dirac operator. The random sources we used are randomly distributed over time, diagonal in spin and dense in the color.}
\label{tab:simudetails}
\end{table} 
In what follows, we only consider the case in which the initial meson is at rest, i.e. we always work in the decaying hadron's reference frame. Moreover, we choose the virtual photon three-momentum  to be in the lattice $z$-direction, namely $\bs{k}=(0,0,k_z)$.  For each value of the photon momentum, the Euclidean correlator in Eq.~(\ref{eq:def_Cmunu_2}) can be determined from the following Euclidean three-point function\footnote{We compute here only the quark-line connected part of the correlation function in Eq.~(\ref{eq:M_munu}), and thus neglect the contribution from the disconnected diagram, which vanishes in the limit of exact $\rm{SU}(3)$ flavour symmetry (see Ref.~\cite{Desiderio:2020oej} for more details). } computed on a finite $L^{3}\times T$ lattice (see Fig.~\ref{fig:Feyn} for a diagrammatic representation)
\begin{align}
M^{\mu\nu}_{W}(t,t_{W},\bs{k}) \equiv \langle J_{\rm em}^{\mu}(t+t_{W},\bs{k}) ~J_{W}^{\nu}(t_{W}) ~\phi^{\dag}_{D_{s}}(0)\rangle_{LT},\qquad t > 0~,
\label{eq:M_munu}
\end{align}
where $\phi^{\dag}_{D_{s}}(0)$ is an interpolating operator, with vanishing three-momentum and located at Euclidean time zero, having the same quantum number of the $D_{s}$ meson, while $t_{W}$ is the fixed time where the weak current $J_{W}^{\nu}$ is inserted, which must be chosen  large enough to ensure the dominance of the ground state\footnote{We have chosen two different values of $t_{W} \simeq 1.75, 2.0~{\rm fm}$, and checked the stability of the results with respect to these two choices. All the following results correspond to the choice  $t_{W} \simeq 1.75~{\rm fm}$.}. The e.m. current inserted at time $t+t_W$ carries three-momentum $\bs{k}$. In the limit of large $t_{W}$ one has
\begin{align}
\label{eq:M_muni_asympt}
M^{\mu\nu}_{W}(t,t_{W},\bs{k}) = \frac{\langle D_{s}(0) |   \phi^{\dag}_{D_{s}} | 0 \rangle}{2M_{D_{s}}}e^{-M_{D_{s}}t_{W}}\cdot \left( C^{\mu\nu}_{W,E}(t,\bs{k})  + \ldots \right)~, 
\end{align}
where the dots represent terms that are exponentially suppressed at large $t_{W}$, and $C^{\mu\nu}_{W,E}(t,\bs{k})$ is the Euclidean correlator, which is related to the spectral density $\rho^{\mu\nu}_{W}(E',\bs{k})$ in Eq.~(\ref{spectralf}) through the relation
\begin{align}
C^{\mu\nu}_{W,E}(t,\bs{k}) = \int_{E^{*}}^{\infty} \frac{dE'}{2\pi} e^{-E't}\rho^{\mu\nu}_{W}(E',\bs{k})~,    
\end{align}
(see Eq.~(\ref{CtE})). Starting from Eq.~(\ref{eq:M_muni_asympt}), 
 the Euclidean correlator is determined exploiting the fact that the prefactor multiplying $C_{W,E}^{\mu\nu}(t,\bs{k})$ in Eq.~(\ref{eq:M_muni_asympt}) can be computed from the knowledge of the two-point function of the $D_{s}$ meson (see Refs.~\cite{Desiderio:2020oej,Gagliardi:2022szw} for more details on this point).

We considered three different values of the photon spatial momentum, $|k_{z}|/M_{D_{s}} \simeq 0.1, 0.25, 0.35$. These momenta, which are not integer multiple of $2\pi/L$, have been obtained imposing twisted boundary conditions for the strange and charm (valence) quark fields in the $\hat{z}$ direction, as explained in Ref.~\cite{Desiderio:2020oej}. We averaged $C^{\mu\nu}_{W,E}$ between opposite photon momenta $\pm k_{z}$, exploiting the symmetry properties of the three-point correlation function in Eq.~(\ref{eq:M_munu})  under $\bs{k} \to -\bs{k}$. The (correlated) average between opposite momenta turns out to be beneficial in reducing the gauge noise at small values of the momentum, particularly in the vector channel. We use an $\mathcal{O}(a)$ improved mixed action lattice setup based on maximally twisted clover Wilson 
quarks~\cite{Frezzotti:2004wz}, for which the details of the $N_f=2+1+1$ sea quark sector are given
in Ref.~\cite{Alexandrou:2018egz}. In the valence we employ Osterwalder-Seiler quarks with flavour diagonal action, choose $r_c = - r_s = \pm 1$ for the charm and strange Wilson parameters, and
adopt a point-like discretization of the weak and electromagnetic current 
\begin{align}
J_W^\nu(t,\bs{x}) &= J_V^\nu(t,\bs{x})-J_A^\nu(t,\bs{x})=Z_{A}\bar{s}(t,\bs{x})\gamma^\nu c(t,\bs{x})
-Z_{V}\bar{s}(t,\bs{x})\gamma^\nu\gamma_5 c(t,\bs{x})\, ,  \label{eq:jWVA}~ \\[11pt]
J_{\rm em}^{\mu}(t,\bs{x}) &= \frac{2}{3}Z_{V}\bar{c}(t,\bs{x})\gamma^\mu c(t,\bs{x}) -\frac{1}{3}
Z_{V}\bar{s}(t,\bs{x})\gamma^\mu s(t,\bs{x}) ~, \label{eq:jem}
\end{align}
and we have dropped from 
 $J^{\mu}_{\rm em}$ the contributions of flavours other than charm and strange, as they do not contribute to the connected part of the correlation function in Eq.~(\ref{eq:M_munu}).
Note that with twisted-mass fermions at maximal-twist and $r_c=-r_s$, the Renormalization Constants (RCs) to be used for the flavoured (cs) currents $J_{V}^{\nu}$ and $J_{A}^{\nu}$ in Eq.~(\ref{eq:jWVA}) are chirally-rotated w.r.t. the ones of standard Wilson fermions, and the vector ($J_{V}^{\nu}$) and axial-vector ($J_{A}^{\nu}$) currents renormalize respectively with the multiplicative RCs $Z_A$ and $Z_V$. The flavour diagonal components of the e.m. current in Eq.~(\ref{eq:jem}) are instead renormalized by $Z_V$.
For the ensemble used in the present study the two RCs are given by $Z_{V} = 0.70638(2)$ and $Z_{A}= 0.7428(3)$~\cite{Alexandrou:2022amy}.    

The spectral density $\rho^{\mu\nu}_{W}$ in Eq.~(\ref{spectralf}) can be separated into two contributions, $\rho^{\mu\nu; s}_{W}$ and $\rho^{\mu\nu;c}_{W}$, corresponding to the emission of the virtual photon from the strange or from the charm quark-line. The two contributions can be easily obtained by plugging in Eq.~(\ref{spectralf}) the $\bar{s}\gamma^{\mu}s$ or the $\bar{c}\gamma^{\mu}c$ part of the electromagnetic current in Eq.~(\ref{eq:jem}) in place of the total $J^{\mu}_{\rm em}$.
The intermediate states contributing to $\rho^{\mu\nu; c}_{W}$ are vector charmonium resonances, the lightest of which is the $J/\Psi$ resonance ($M_{J/\Psi} \sim 3.1~{\rm GeV}$), while those contributing to $\rho^{\mu\nu; s}_{W}$ are vector $s\bar{s}$ states, the lightest of which is the $\phi$ meson, which then decays via strong interactions mainly into $K^{+}K^{-}$ states. In the $D_{s}$ meson reference frame, the largest possible photon energy $E$ allowed in the $D_{s}\to \ell\nu\gamma^{*}$ decay, is clearly given by $M_{D_{s}}$ (neglecting the small lepton mass). Since $M_{J/\Psi} > M_{D_{s}} > E$, the charm contribution to the hadronic tensor $H^{\mu\nu}_{W,2}(E,\bs{k})$ can be computed from the knowledge of the corresponding Euclidean correlator using the standard approach based on Eq.~(\ref{ampE}), without encountering the problems related to the analytic continuation. Instead, the spectral density for the strange component, $\rho^{\mu\nu; s}_{W}(E', \bs{k})$, becomes non-zero at the two-kaon threshold $E^{*} = \sqrt{ 4M_{K}^{2} + |\bs{k}|^{2}}$, in the infinite-volume limit. However, with the typical spatial volumes adopted in numerical simulations, including the one used in the present work, the smallest relative momentum than can be carried by the $K^{+}K^{-}$ pair (which must be in a $J=1$ state) is such that its energy is larger than the $\phi$-resonance mass $M_{\phi} \sim 1019~{\rm MeV}$, which is thus effectively a stable hadron. In this case one clearly has $E^{*} = \sqrt{ M_{\phi}^{2} + |\bs{k}|^{2}}$, and for $E > E^{*}$ we can use our approach based on the calculation of the smeared amplitudes. Since the charm-quark contribution can be computed using the standard approach, in the following we will discuss the results corresponding to the strange-quark contribution only. To keep the notation compact, however, we will continue to use $\rho^{\mu\nu}_{W}$, $C^{\mu\nu}_{W,E}$ without the superscript ``s", although it is understood that in all the following results the charm-quark contribution has been omitted.

\subsection{The HLT method}
\label{sec:HLT}
The general expression in Eq.\,(\ref{smH}) for the hadronic amplitude allows for the application of the HLT method developed in Ref.~\cite{Hansen:2019idp}  for the evaluation of the smeared amplitudes (see also Refs.~\cite{Alexandrou:2022tyn, Evangelista:2023vtl} for two recent applications). In this subsection, we summarize the main ingredients of the procedure. The goal is to find, for non-zero value of the smearing parameter $\varepsilon$, the best approximation of the kernel $K(E'-E;\varepsilon)$ entering the representation of the smeared hadronic tensor\footnote{With respect to Eq.~(\ref{smH}) we have dropped the suffix $K$ from the definition of the smeared hadronic tensor to lighten the notation. In what follows all the results have been obtained using the lattice kernel in Eq.~(\ref{Cauchylat}).}

\begin{align}
 H^{\mu\nu}_{W,2}(E, \bs{k};\varepsilon) \equiv \int_{E^{*}}^{\infty} \frac{ dE'}{2\pi} K(E'-E;\varepsilon) \rho^{\mu\nu}_{W}(E',\bs{k})~, \qquad \lim_{\varepsilon\to 0} H^{\mu\nu}_{W,2}(E,\bs{k};\varepsilon) = H^{\mu\nu}_{W,2}(E,\bs{k})~, 
\end{align}
in terms of the basis function $\{ e^{-aE' n}
\}_{n=1,\ldots, n_{max} }$, namely
\begin{flalign}
\label{eq:basis_expansion}
{\rm Re} [K(E'-E;\varepsilon) ] \simeq \sum_{n=1}^{n_{max}} g^{R}_{n}(E,\varepsilon) e^{-aE' n}~,\qquad {\rm Im}[K(E'-E;\varepsilon)] \simeq \sum_{n=1}^{n_{max}} g^{I}_{n}(E,\varepsilon) e^{-aE' n}~,
\end{flalign} 
where $a$ is the lattice spacing and the dimension of the exponential basis $n_{max}$ is typically chosen to be equal to the number of discrete lattice times at which the Euclidean correlator $C^{\mu\nu}_{W,E}(t,\bs{k})$ is known\footnote{Due to the periodicity of the lattice in the temporal direction, for $t+t_{W}>T/2$ the around-the-world contributions from time orderings different than the one considered, contaminate our determination of $C^{\mu\nu}_{W,E}(t,\bs{k})$. For this reason, we do not consider times $t> T/2-t_{W}$, where the electromagnetic current is placed in the second half of the lattice.}, in our case $n_{max} = (T/2
-t_{W})/a
$. In this way,  once the coefficients $g_{n}^{R}$ and $g_{n}^{I}$ are known, the smeared hadronic tensor $H^{\mu\nu}_{W,2}(E,\bs{k};\varepsilon)$ can be obtained, from the knowledge of $C^{\mu\nu}_{W,E}(t,\bs{k})$ only, via
\begin{align}
\label{eq:HLT_RE_IM}
H^{\mu\nu}_{W,2}(E,\bs{k};\varepsilon) &= \int_{E^{*}}^{\infty} \frac{dE'}{2\pi} \left( K_{R}(E'-E;\varepsilon) + i K_{I}(E'-E;\varepsilon)\right)\rho^{\mu\nu}_{W}(E',\bs{k}) \nonumber \\[10pt]
&\simeq \sum_{n=1}^{n_{max}} \left( g_{n}^{R}(E,\varepsilon) +i g_{n}^{I}(E,\varepsilon)\right) \int_{E^{*}}^{\infty} \frac{dE'}{2\pi} e^{-aE' n} \rho^{\mu\nu}_{W}(E',\bs{k}) 
= \sum_{n=1}^{n_{max}} \left( g_{n}^{R}(E,\varepsilon) + ig_{n}^{I}(E,\varepsilon)\right) ~ C^{\mu\nu}_{W,E}(an, \bs{k})~,
\end{align}
where $K_{R/I}$ are the real/imaginary part of the kernel function.
As discussed thoroughly in Ref.~\cite{Hansen:2019idp}, the problem of finding the coefficients $g^{R/I}_{n}$ presents a certain number of technical difficulties. Any determination of the smeared hadronic tensor $H^{\mu\nu}_{W,2}(E,\bs{k};\varepsilon)$ based on Eqs.~(\ref{eq:basis_expansion}),~(\ref{eq:HLT_RE_IM}) will be inevitably affected by both systematic errors (due to the inexact reconstruction of the kernels) and statistical uncertainties (due to the fluctuations of the correlator $C^{\mu\nu}_{W,E}$), which need to be simultaneously kept under control. If we were only concerned with systematic errors, the best coefficient $g_{n}^{R/I}$ could be obtained by minimizing the quadratic form\footnote{ In Eq.~(\ref{eq:func_A}), $E_{th}<E^{*}$ is an adjustable algorithmic parameter. In all the results that will be shown its value has been set to $E_{th}=0.9E^{*}$. However, we checked the stability of the results by repeating the analysis using different values of $E_{th}$.}
\begin{align}
\label{eq:func_A}
A_{R/I}[\bs{g}]=
\int_{E_{th}}^\infty\dd E' \,  \abs{ \sum_{n=1}^{n_{max}}~g_{n} e^{-aE' n} - K_{R/I}(E'-E;\varepsilon)}^2~, \quad \bs{g}= (g_{1},\ldots,g_{n_{max}}), \qquad E_{th} \leq E^{*}~.   
\end{align}
However, for small values of $\varepsilon$, the coefficients resulting from the minimization of $A_{R/I}[\bs{g}]$ turn out to be very large in magnitude and oscillating in sign, strongly amplifying the statistical errors of the correlator $C^{\mu\nu}_{W,E}$ when the smeared hadronic tensor is evaluated using Eq.~(\ref{eq:HLT_RE_IM}).

\begin{figure}
    \centering
\includegraphics[scale=0.4]{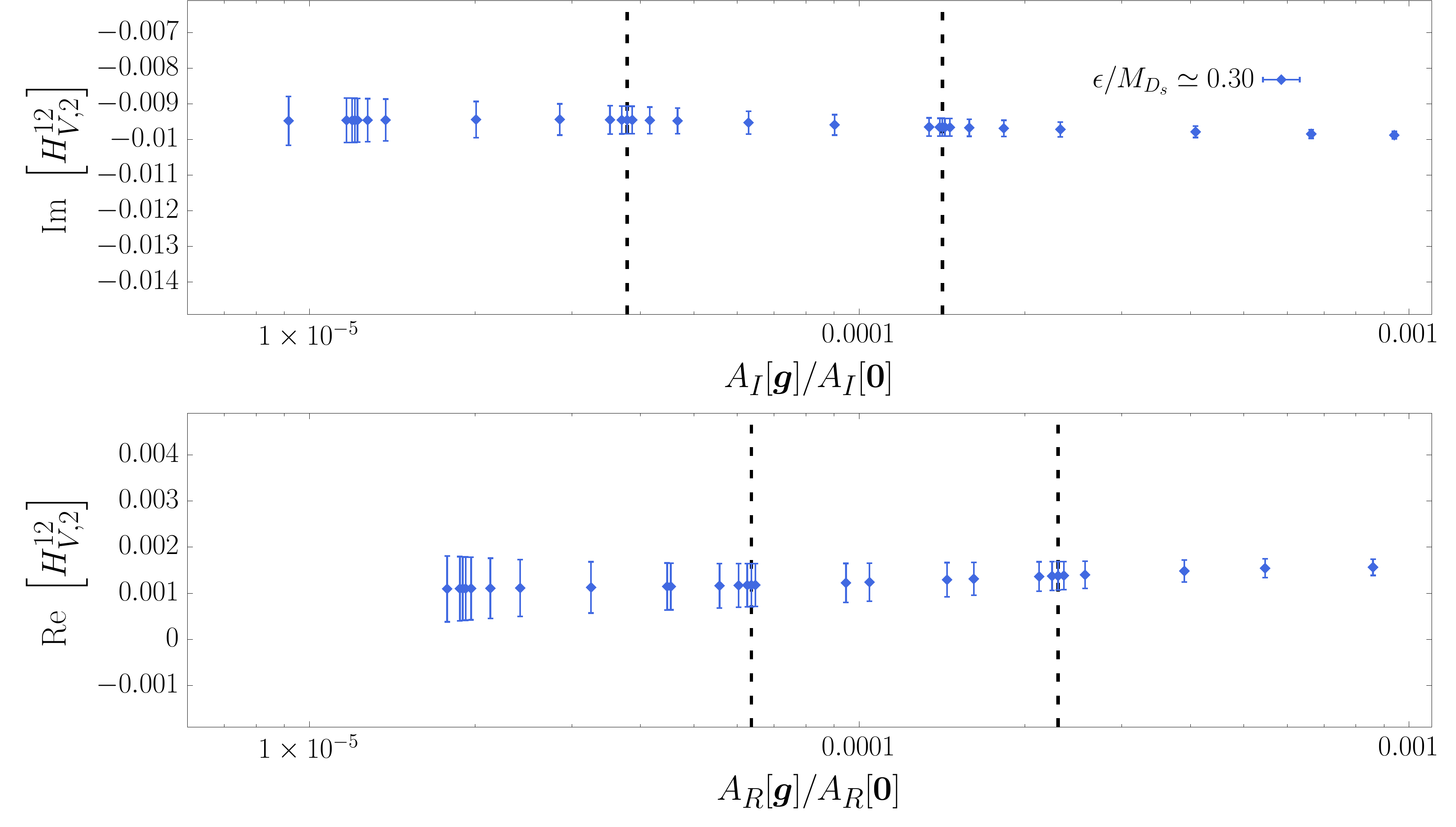}
    \caption{\it\small  The real (bottom figure) and imaginary (top figure) part of $H^{12}_{V}(E,\bs{k};\varepsilon)$ in lattice units, for $\varepsilon \simeq 0.3M_{D_{s}}$, photon momentum $|\bs{k}|\simeq 0.1M_{D_{s}}$, and fixed photon energy $E \simeq 1.2~{\rm GeV}$, as a function of the ratio $A_{R/I}[\bs{g}]/A_{R/I}[\bs{0}]$ indicating the quality of the kernel reconstruction obtained employing different values of $\lambda$. The plot shows an example of our stability analysis. The rightmost and leftmost vertical lines correspond to the reconstructions obtained using $\lambda=\lambda^{opt.}$ and $\lambda=\lambda^{syst.}$, respectively, see text for details.}
    \label{fig:stab_analysis}
\end{figure}

The HLT method developed in Ref.~\cite{Hansen:2019idp}, provides a regularization mechanism to this problem, enabling to find an optimal balance between statistical and systematic errors. This is achieved by minimizing a linear combination
\begin{align}
\label{eq:func_W}
W_{R/I}[\boldsymbol{g}] \equiv \frac{A_{R/I}[\boldsymbol{g}]}{A_{R/I}[\boldsymbol{0}]} + \lambda B[\boldsymbol{g}]\;,
\end{align}
of the norm-functional $A_{R/I}[\boldsymbol{g}]$ of Eq.~(\ref{eq:func_A}) and of the error-functional
\begin{align}
B[\boldsymbol{g}] =  B_{\rm{norm}}\sum_{n_1 , n_2 = 1}^{n_{max}} g_{n_{1}}\, g_{n_{2}}~ {\rm{Cov}}(an_1 , an_2 )~, 
\end{align}
where ${\rm Cov}$ is the covariance matix of the Euclidean lattice correlator $C^{\mu\nu}_{W,E}$, and $\lambda$ is the so-called \textit{trade-off} parameter. $B_{\rm norm}$ is a normalization factor introduced to render dimensionless the parameter $\lambda$. In the absence of statistical errors, the functional in Eq.~(\ref{eq:func_W}) reduces to that in Eq.~(\ref{eq:func_A}), up to an irrelevant multiplicative factor. Instead, in the presence of statistical errors, the functional $B$ disfavours coefficients $\bs{g}$ leading to too large statistical uncertainties in the reconstructed hadronic amplitude. The balance between having small systematic errors (small $A_{R/I}[\bs{g}]$) and small statistical errors (small $B[\bs{g}]$) depends on the tunable parameter $\lambda$. Its optimal value $\lambda^{opt.}$ is determined by monitoring the reconstructed smeared hadronic tensor for different $\lambda$. The optimal value is then chosen in the statistically-dominated regime, where $\lambda$ is sufficiently small that the systematic error due to the kernel reconstruction is smaller than the statistical error (therefore, in this region, the results are stable under variations of $\lambda$ within statistical uncertainties), but large enough to still have reasonable statistical errors. An illustrative example of this \textit{stability analysis} is given in Fig.~\ref{fig:stab_analysis} in the case $W=V$ and for a fixed photon energy $E> E^{*}$, where in our case $E^{*}=\sqrt{ M_{\phi}^{2} + |\bs{k}|^{2}}$. The rightmost vertical line appearing in each of the two plots of Fig.~\ref{fig:stab_analysis}, corresponds to the chosen optimal value $\lambda^{opt.}$, while the rightmost corresponds to the value $\lambda^{syst.}$ determined imposing
\begin{align}
\frac{B[\bs{g}_{\lambda^{syst.}}]}{A_{R/I}[ \bs{g}_{\lambda^{syst}}]} = \kappa\frac{B[\bs{g}_{\lambda^{opt.}}]}{A_{R/I}[ \bs{g}_{\lambda^{opt.}}]}~,
\end{align}
and we choose $\kappa = 10$.
The difference between the reconstructions obtained using $\lambda=\lambda^{opt.}$ and $\lambda=\lambda^{syst.}$ is added as a systematic uncertainty in the final error (see the Supplementary Material of Ref.~\cite{Alexandrou:2022tyn} for more details on this point). 

\subsection{Smeared hadronic amplitudes from the HLT method}
We now present our determination of the real and imaginary part of the smeared hadronic tensor $H^{\mu\nu}_{W;2}(E,\bs{k};\varepsilon)$. As discussed in sect.~\ref{sec:HLT}, the statistical uncertainties increase for decreasing values of the smearing parameter $\varepsilon$, and with the current statistical precision of the Euclidean correlator the smallest value of $\varepsilon$ we can reach for $E > E^{*}$ is $\varepsilon \simeq 100~{\rm MeV}$. Since in this work we are mainly interested to study the effectiveness of the method, rather than giving results of immediate phenomenological application (which will be part of future studies), we discuss here only the results relative to the vector channel $W=V$. We mention however that the analysis and the results in the axial channel display the same qualitative features as the vector case. For the same reason, all the results shown correspond to a single value of the  photon three-momentum $|k_{z}| = M_{D_{s}}/10$, since the same qualitative behaviour has been observed for all the other simulated photon momenta. 

With the photon momentum directed along the $z-$axis, only the component $H^{12}_{V,2} = - H^{21}_{V,2}$ is different from zero, and is proportional to the vector form factor $F_{V}$~(see e.g. Refs.~\cite{Desiderio:2020oej,Gagliardi:2022szw} for the full form factor decomposition of $H^{\mu\nu}_{W})$. To keep a reader-friendly notation, and in analogy with what has been done in sect.~\ref{sec:method}, in the following we will indicate $C^{12}_{V,E}(t,\bs{k})$ with $C_{E}(t)$, $H^{12}_{V,2}(E,\bs{k};\varepsilon)$ with $H(E;\varepsilon)$,  $H^{12}_{V,2}(E,\bs{k})$ with $H(E)$,  and $\rho^{12}_{V}(E,\bs{k})$ with $\rho(E)$.  

\begin{figure}
    \centering
    \includegraphics[scale=0.29]{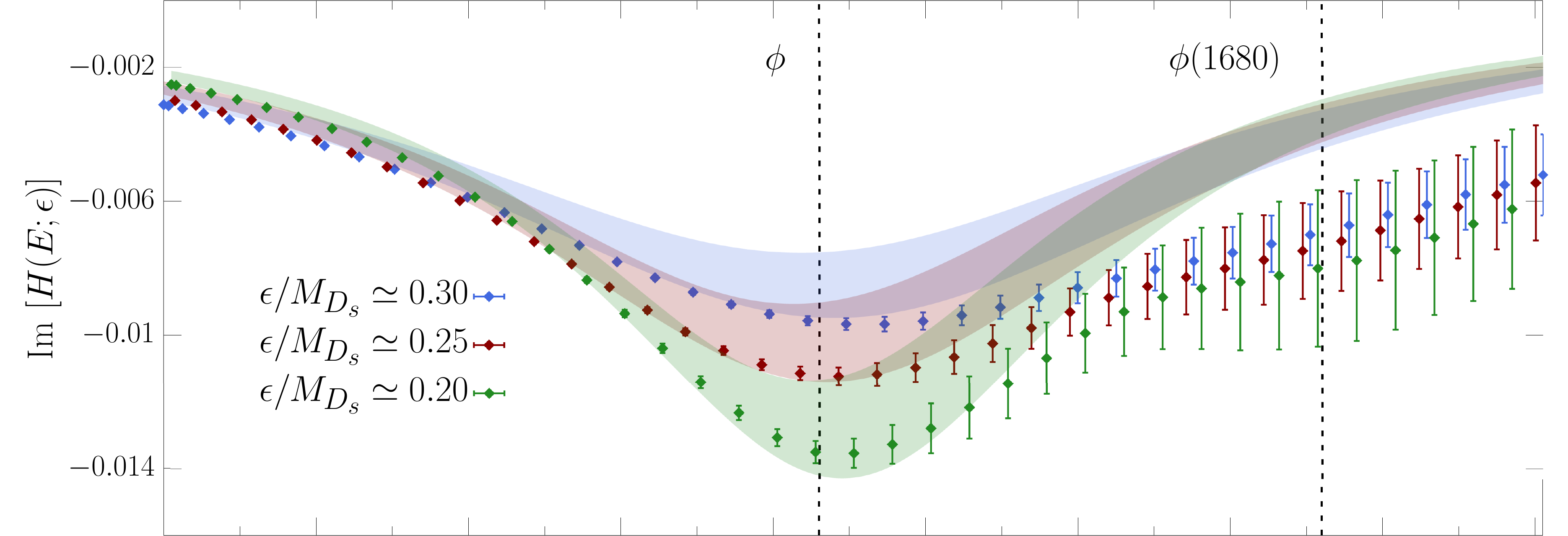}
    \centering
    \includegraphics[scale=0.29]{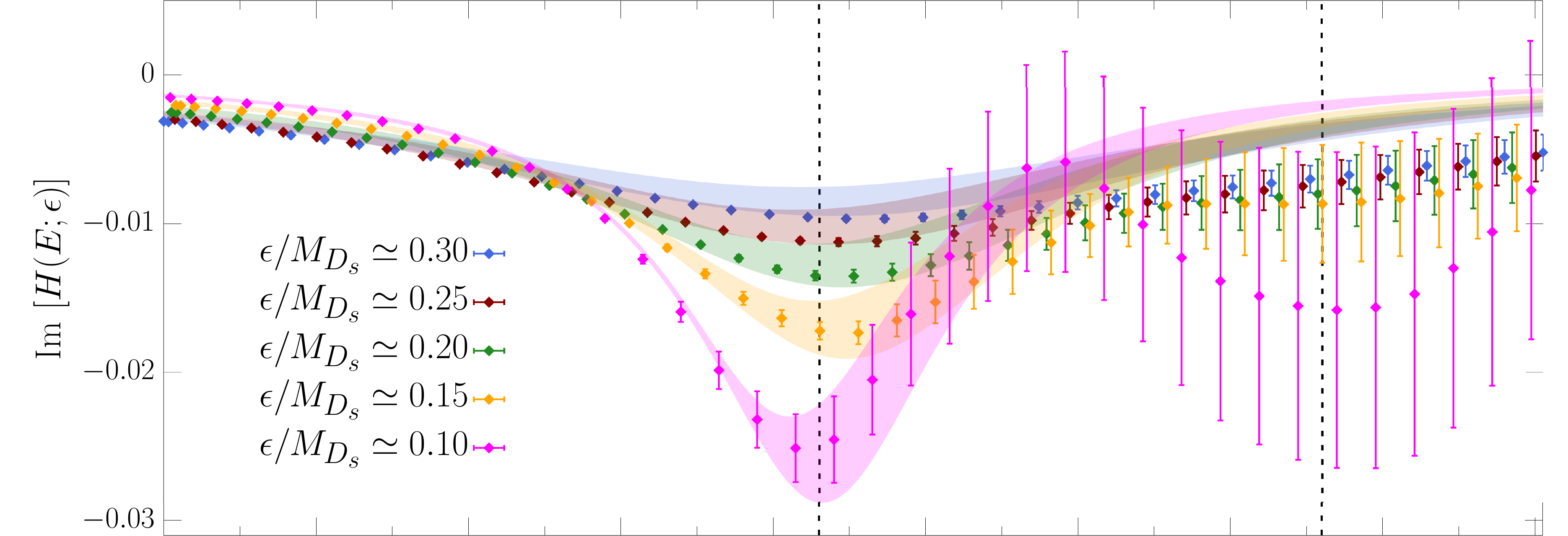}
    \centering
    \includegraphics[scale=0.29]{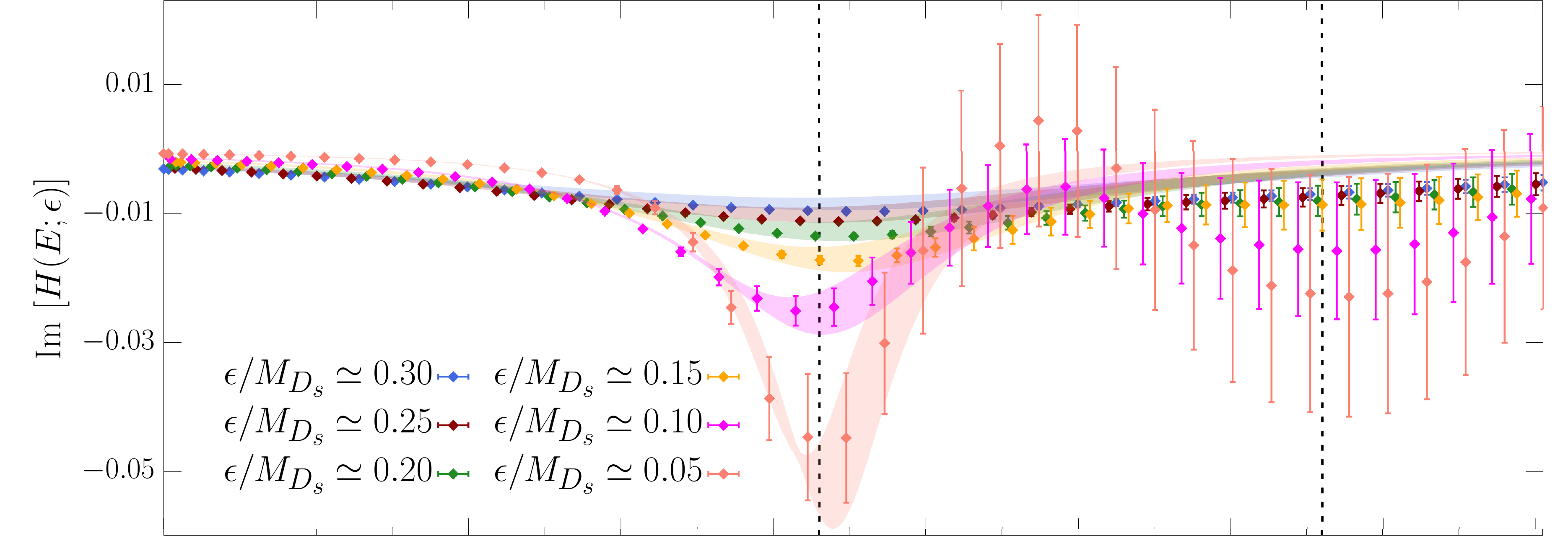}\\[2pt]
    \centering
    \includegraphics[scale=0.29]{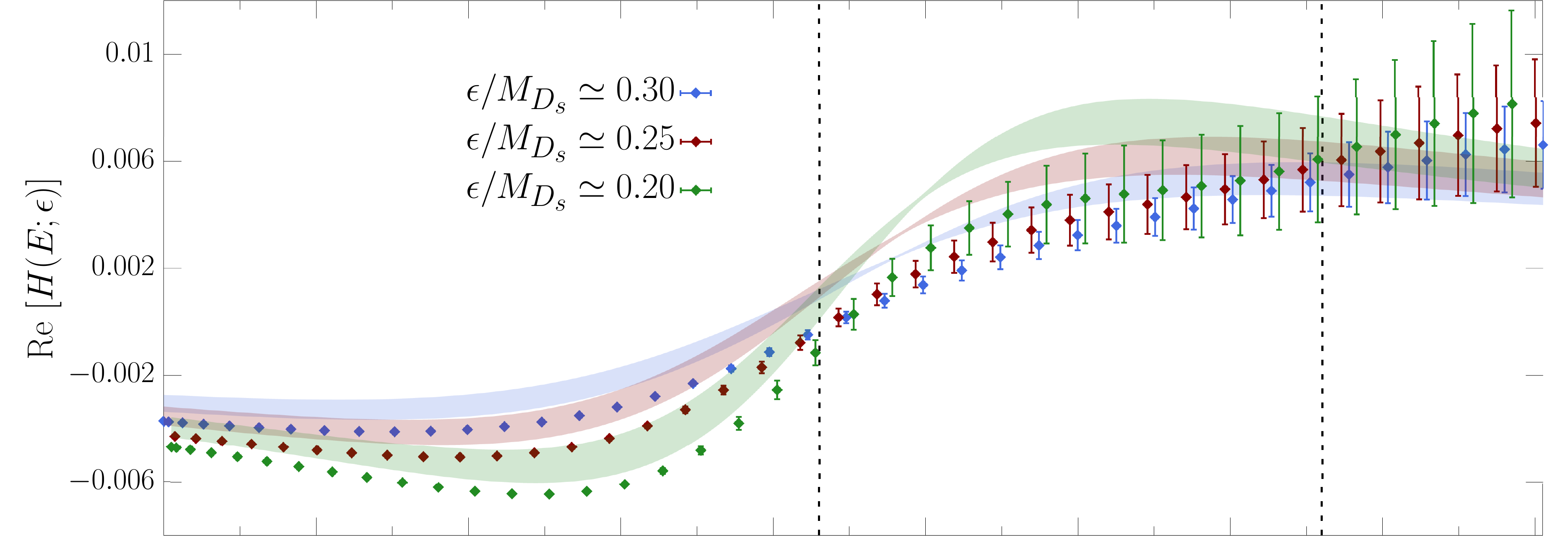}
    \centering
    \includegraphics[scale=0.29]{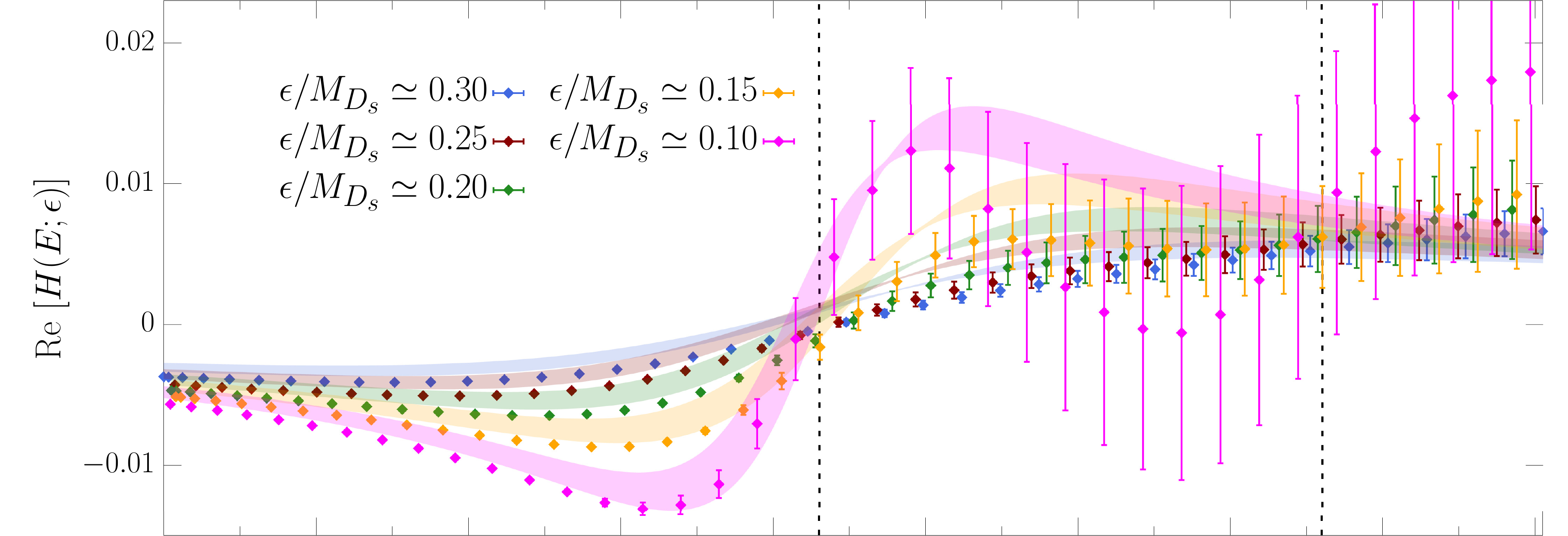}
    \centering
    \includegraphics[scale=0.29]{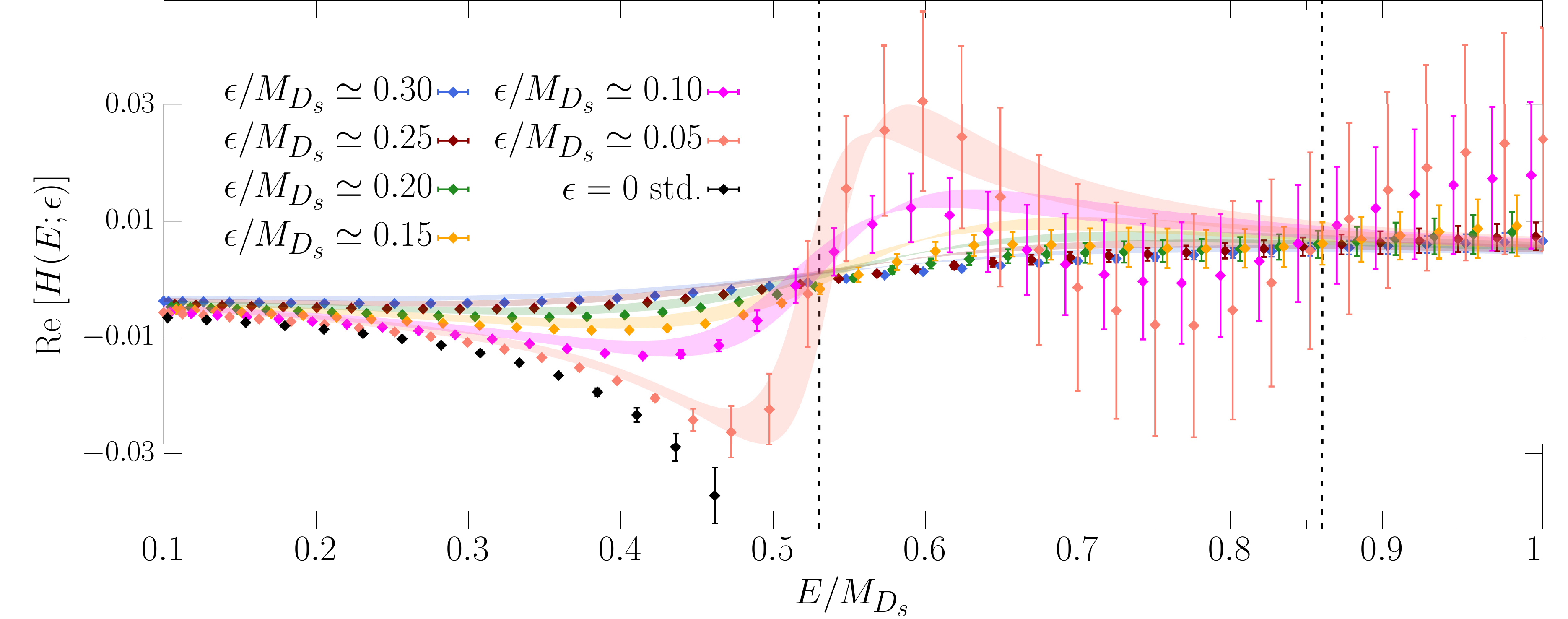}

    \caption{\it\small  The real (bottom three plots) and imaginary (top three plots) part of $H(E, \varepsilon)$ in lattice units for different values of $\varepsilon/M_{D_{s}}$, as a function of $E/M_{D_{s}}$. In each plot, the coloured bands correspond to the prediction of the VMD model of Eq.~(\ref{eq:VMD}). The two vertical dashed lines correspond to the position of the $\phi(1019)$ and $\phi(1680)$ resonances. The black data points labelled with $\varepsilon=0$ std. correspond to the determination of ${\rm Re}[H(E)]$ obtained using the standard approach based on Eq.~(\ref{ampE}). }
    \label{fig:E_dep}
\end{figure}

In Fig.~\ref{fig:E_dep} we show our determination of the real (bottom three plots) and imaginary (top three plots) part of $H(E;\varepsilon)$ as a function of the dimensionless parameter $E/M_{D_{s}}$, for different values of the smearing parameter $\varepsilon$ in the energy range $\varepsilon \in [0.1, 0.6]~{\rm GeV}$. For both ${\rm Re}[H(E;\varepsilon)]$ and ${\rm Im}[H(E;\varepsilon)]$, from top to bottom, we show results at progressively smaller values of the smearing parameter $\varepsilon$. The third and last plots contain, for ${\rm Im}[H(E;\varepsilon)]$ and ${\rm Re}[H(E;\varepsilon)]$ respectively, the results on the entire range of $\varepsilon$ we explored, with the exception of $\varepsilon= 50~{\rm MeV}$ which for $E > E^{*} \simeq 0.53 M_{D_{s}}$ turns out to have too large statistical errors. The uncertainties shown in all the plots include the systematic errors due to the kernel reconstruction, estimated following the procedure described in the previous section.

As it is clear from the plots of Fig.~\ref{fig:E_dep}, the reconstructed smeared amplitudes display the same qualitative features observed in the model of Fig.~\ref{fig:smH-model}, with the imaginary part showing a clear peak, and the real part the typical behaviour of the regularized principal part of $1/(E-E^{*})$, in correspondence of the energy $E^{*}$ of the $\phi$ meson. In the lowest plot of Fig.~\ref{fig:E_dep}, we also show for comparison the determination of ${\rm Re}[H(E)]$ obtained using the standard approach based on Eq.~(\ref{ampE}), which as already discussed, it is equivalent to apply the spectral method directly with $\varepsilon = 0$ and can be used only for $E < E^{*}$. As the figure shows, for $E<E^{*}$, our data for ${\rm Re}[H(E;\varepsilon)]$ get closer and closer to the result of the standard approach as the smearing size is reduced. This point will be further discussed in more details in the next subsection. 

In each of the plots of Fig.~\ref{fig:E_dep} the transparent bands correspond to the vector-meson dominance (VMD) prediction, where one assumes the dominance of the contribution from the lightest intermediate state. This is obtained by approximating the Euclidean correlator $C_{E}(t)$ with its lowest-lying exponential, namely 
\begin{equation}
\label{eq:VMD}
 C_{E}(t) \simeq A\,e^{ -E_{0}t}~,
\end{equation}
and the spectral density with a single delta function
\begin{equation}
\rho(E) = 2\pi A~\delta (E-E_{0}) ~,   
\end{equation}
where in our case $E_{0}$ corresponds to the energy of the $\phi$ meson. The VMD approximation simply corresponds to the one-resonance model in Eq.~(\ref{eq:mod}) with $M=E_{0}$, in the limiting case $\Gamma = 0$. Both the amplitude $A$ and the energy $E_{0}$, are determined through a standard effective mass/residue analysis of the lattice correlator $C_{E}(t)$. We observe that the VMD model works rather well, reproducing the main qualitative features of the full result. However, important differences can be appreciated as well. For the real part, we observe that the lattice data up to $E \simeq 0.6 M_{D_{s}}$ are shifted downwards w.r.t. the VMD prediction. This shift can be understood assuming that all intermediate states $| n \rangle$ but the lightest, have energies $E_{n} \gg E$. In this case, in the limit $\varepsilon =0$ one has\footnote{ In writing down Eq.~(\ref{eq:spec_decomposition}) we are assuming to be in a finite volume where the spectral density is always a sum of isolated delta peaks. This is not the case in the infinite-volume limit due to the presence of continuum multi-particle states.}
\begin{equation}
\label{eq:spec_decomposition}
{\rm Re}[H(E<E_{0})] = \frac{A}{E_{0} - E}   + \sum_{n = 1}^{\infty} \frac{A_{n}}{E_{n} - E} \simeq  
\frac{A}{E_{0}- E  } + \sum_{n=1}^{\infty} \frac{A_{n}}{E_{n}} =  \frac{A}{E_{0}- E  } + {\rm const.}~, 
\end{equation}
where
$
A_{n}  \equiv \langle 0 | J_{{\rm em}}^{1} | n \rangle \langle n | J_{V}^{2} | D_{s} \rangle  /2E_{n}$. Therefore, for $E\ll E_{n}$, the dependence on the energy $E$ of the heavier intermediate states contributions can be neglected, and the full result appears to be approximately shifted by a constant w.r.t. the pure VMD prediction. 
 
The presence of contributions from intermediate states other than the $\phi$ meson, can also be observed in the plots of the imaginary part of $H(E;\varepsilon)$, where one expects to observe a peak, smeared over an energy interval $\varepsilon$, in correspondence of any of the energies $E_{n}$ of the intermediate states. From Fig.~\ref{fig:E_dep} it can be noticed that our data for ${\rm Im}[H(E;\varepsilon)]$, unlike the VMD predictions, do not go to zero after the $\phi$-meson peak, thus signalling the presence of intermediate states lying in an interval of $\mathcal{O}(\varepsilon)$ around $E$. In the case at hand, a sizeable contribution to the spectral density is expected in correspondence of the $\phi(1680)$ resonance~\cite{ParticleDataGroup:2020ssz} which has a mass $M_{\phi(1680)} = 1680(20)~{\rm MeV}$, and a total decay width $\Gamma_{\phi(1680)} = 150(50)~{\rm MeV}$. The position of the $\phi(1680)$ is indicated by the rightmost vertical dashed line in the plots of Fig.~\ref{fig:E_dep}. Although the present statistical accuracy of the Euclidean correlator $C_{E}(t)$ does not allow us to determine $H(E;\varepsilon)$ with reasonable errors for $\varepsilon \lesssim 100~{\rm MeV}$, the presence of a second peak in ${\rm Im}[H(E;\varepsilon)]$, in correspondence to the position of the $\phi(1680)$ resonance can be appreciated in the second and third plot of Fig.~\ref{fig:E_dep}. 

Finally, we notice that for $E < E^{*}$, while the imaginary part of $H(E)$ is exactly zero, this is not true for the smeared amplitude $H(E;\varepsilon)$, since  $K_{I}(E'-E;\varepsilon)$ is a smeared delta function at non-vanishing $\varepsilon$. Therefore, the imaginary part of the convolution integral in Eq.~(\ref{master1}) receives contributions from the region $E' > E^{*}$, also for $E < E^{*}$. In the next subsection, we will come back to this point and discuss how vanishing imaginary parts are recovered for $E< E^{*}$ in the limit $\varepsilon\to 0$.

\subsection{The limit of vanishing \texorpdfstring{$\varepsilon$}{epsilon}}
\label{sec:eps_extrapolation}
We now turn into the discussion of the extrapolation of the real and imaginary part of $H(E;\varepsilon)$ to vanishing smearing parameter $\varepsilon$. Let us start from values of $E$ below the threshold $E^{*}$, where we expect to recover the standard result based on Eq.~(\ref{ampE}), with the imaginary part being exactly zero. Based on the analysis of the $\varepsilon$-expansion in Eqs.~(\ref{eq:exp_RE}),~(\ref{eq:exp_IM}), one expects the corrections to the $\varepsilon=0$ limit to be described by a polynomial in $\varepsilon$, containing both even and odd powers. However, as already pointed out in sect.~\ref{sec:epsexpansion}, since for $E < E^{*}$ the spectral density $\rho(E)$ is exactly zero, its derivative $\rho'(E)$ vanishes as well in this region, and for this reason, following Eq.~(\ref{eq:exp_RE}), the expansion of the real part of $H(E;\varepsilon)$ is expected to start at $\mathcal{O}(\varepsilon^{2})$. 

\begin{figure}
    \centering
    \includegraphics[scale=0.45]{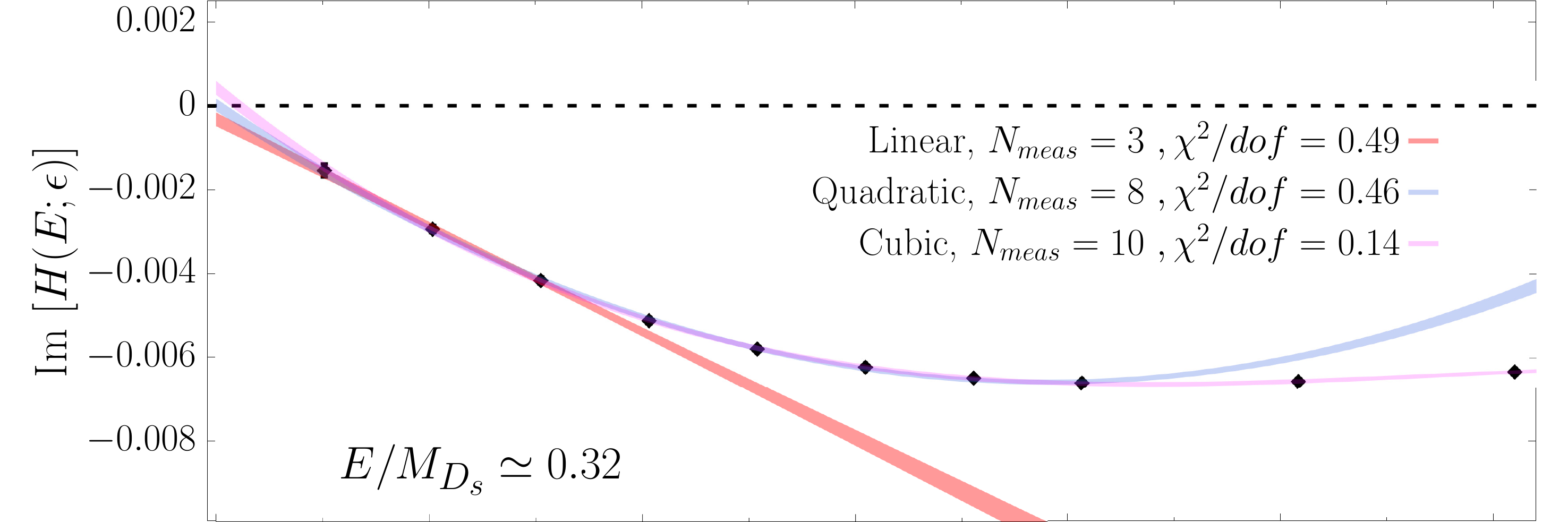}\\
    \centering
    \includegraphics[scale=0.45]{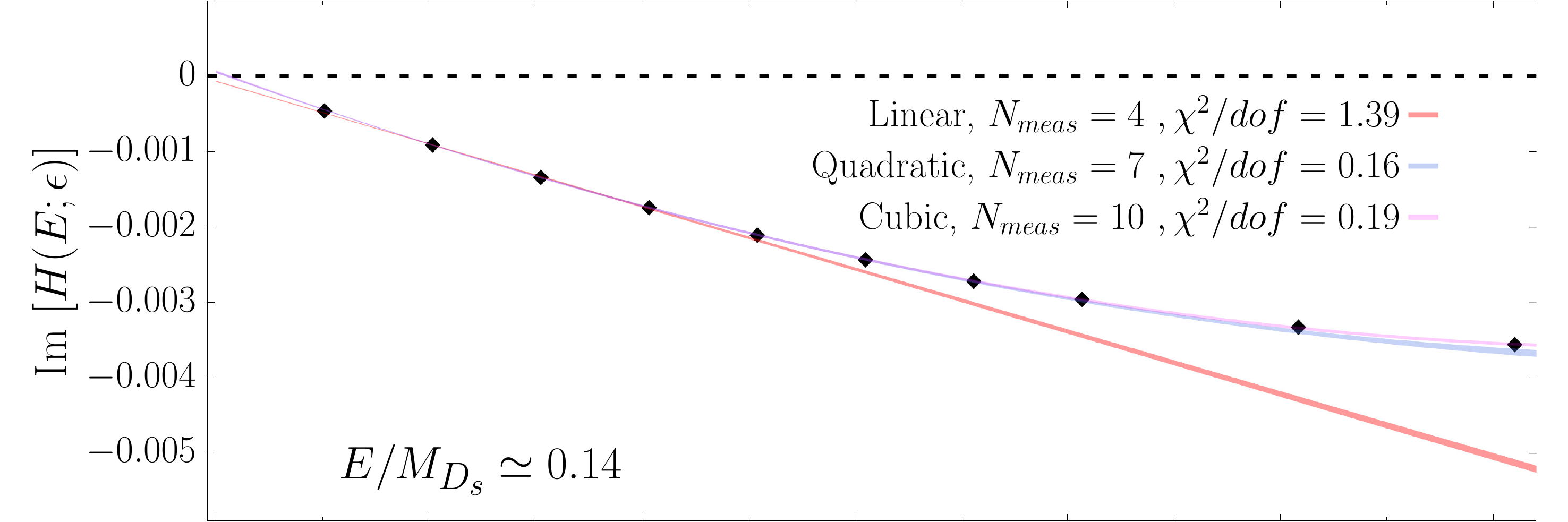}\\[4pt]
    \centering
    \includegraphics[scale=0.45]{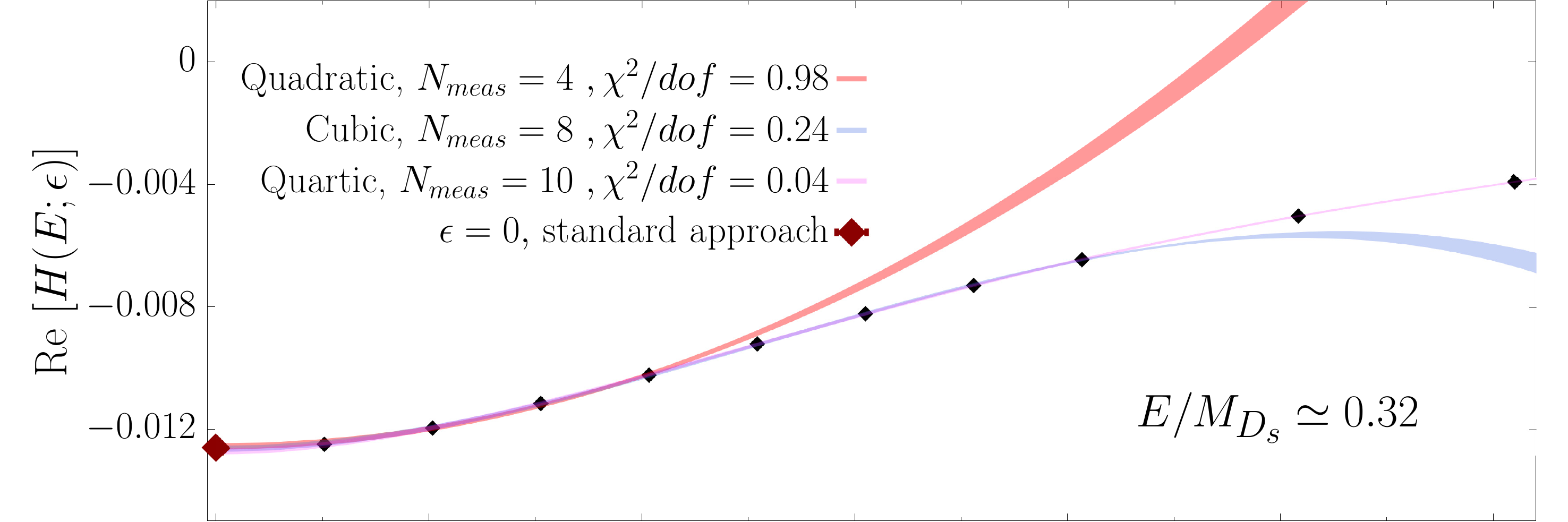}\\
    \centering
    \includegraphics[scale=0.45]{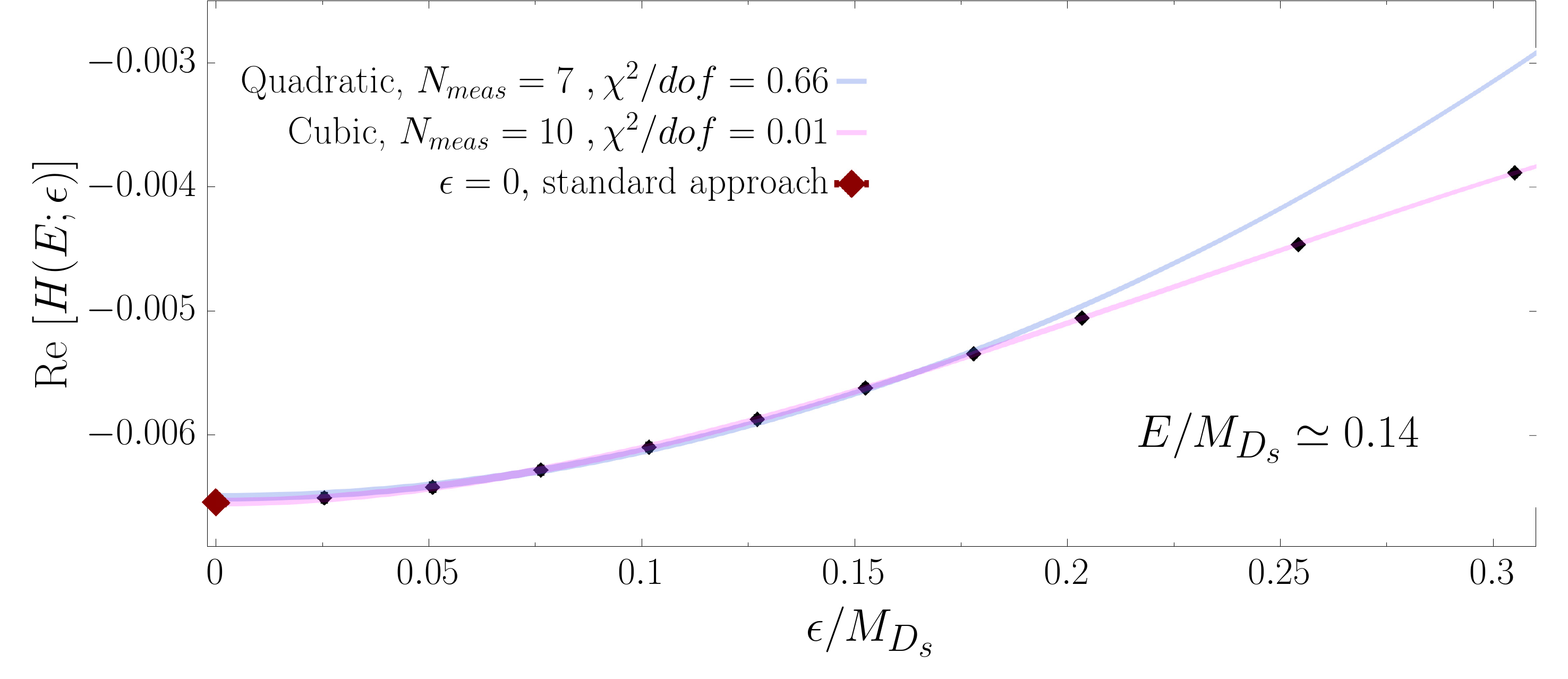}
    \caption{\it\small  Extrapolation to vanishing $\varepsilon$ of the real (bottom two plots) and imaginary (top two plots) part of $H(E;\varepsilon)$ for two fixed values of the dimensionless variable $E/M_{D_{s}}$. The different bands correspond to extrapolations performed employing polynomials in $\varepsilon$ of different degree. $N_{meas}$ is the number of data points used in a given fit, and when $N_{meas}$ $\ne$ $N_{tot}$ $=$ $10$, the $N_{tot}$ $-$ $N_{meas}$ measurements at the largest values of $\varepsilon$ have been excluded. In the fits to ${\rm Re}[H(E;\varepsilon)]$, the fit parameter ${\rm Re}[D_{1}(E)]$ has been fixed to zero. In the bottom two plots the red data points at vanishing $\varepsilon$ corresponds to the result obtained using the standard approach based on Eq.~(\ref{ampE}). Results are given in lattice units.}
    \label{fig:eps_extrapolation}
\end{figure}

In order to extrapolate to vanishing $\varepsilon$, we thus employ the following fourth-order polynomial fit Ansatz
\begin{align}
\label{eq:ansatz}
H(E;\varepsilon) = \bar{H}(E) + D_{1}(E)\varepsilon + D_{2}(E)\varepsilon^{2} +  D_{3}(E)\varepsilon^{3} + D_{4}(E)\varepsilon^{4}  ~,   
\end{align}
where $\bar{H}(E)$ and $\{D_{i}(E)\}_{i=1,\ldots,4}$ are complex-valued free fit parameters. Following the previous discussion, we always set ${\rm Re}[D_{1}(E < E^{*})]=0$. For each value of the photon energy $E$ considered, we performed several fits  by either including or excluding some of the higher order fit coefficients $D_{i}(E)$, or by imposing cuts to the data. The results of the extrapolation are shown in Fig.~\ref{fig:eps_extrapolation}  for two different values of the virtual photon energy $E\simeq 0.14M_{D_{s}}$ and $E\simeq 0.32M_{D_{s}}$.  As the figure shows, below the threshold we are able to recover through the extrapolation the expected results, i.e. vanishing imaginary parts and the standard approach result of Eq.~(\ref{ampE}) for the real part, which is shown as a red data point at $\varepsilon=0$ in Fig.~\ref{fig:eps_extrapolation}. Moreover, the fact that we find no evidence for the $\mathcal{O}(\varepsilon)$ term in the fit to ${\rm Re}[ H(E;\varepsilon)]$ below threshold, and the data approach the $\varepsilon\to 0$ limit quadratically, validates numerically the asymptotic expansion obtained in sect.~\ref{sec:smearedH}.

Around and above the threshold, the extrapolation to vanishing $\varepsilon$ deserves instead more discussion. The analysis of the $\varepsilon-$behaviour of the smeared amplitude $H(E;\varepsilon)$ in the model of Eq.~(\ref{eq:mod}) with a single Breit-Wigner resonance of width $\Gamma$, shows that a polynomial extrapolation can generally be trusted only if the smearing $\varepsilon$ is smaller than $\Gamma$ or if the energy $E$ is sufficiently far from the resonance, i.e. $|E-M| \gg \varepsilon$, with $M$ being the position of the resonance peak. This is expressed by the condition $\varepsilon\ll \Delta(E)$, where $\Delta(E) = \sqrt{(E-M)^2 + \Gamma^2}$. This implies that a polynomial extrapolation in $\varepsilon$, in an energy region close to a very sharp resonance, is generally out of reach, as it requires to compute $H(E;\varepsilon)$ at extremely small values of $\varepsilon$, where the statistical errors are typically too large for the measurement to be of any use. In the case at hand, the spectral density $\rho(E)$ has, in the infinite-volume limit, a sharp peak in correspondence of the $\phi$ resonance which has a decay width $\Gamma_{\phi} \simeq 5~{\rm MeV}$. The smallest amount of smearing $\varepsilon$ we can afford with the present statistical accuracy is instead of order $\mathcal{O}(100~{\rm MeV})$. This implies that in a region of size $100-200~{\rm MeV}$ around $E^{*} \simeq 0.53 M_{D_{s}}$, a polynomial $\varepsilon\to 0$ extrapolation cannot be performed. 

\begin{figure}
    \centering
    \includegraphics[scale=0.51]{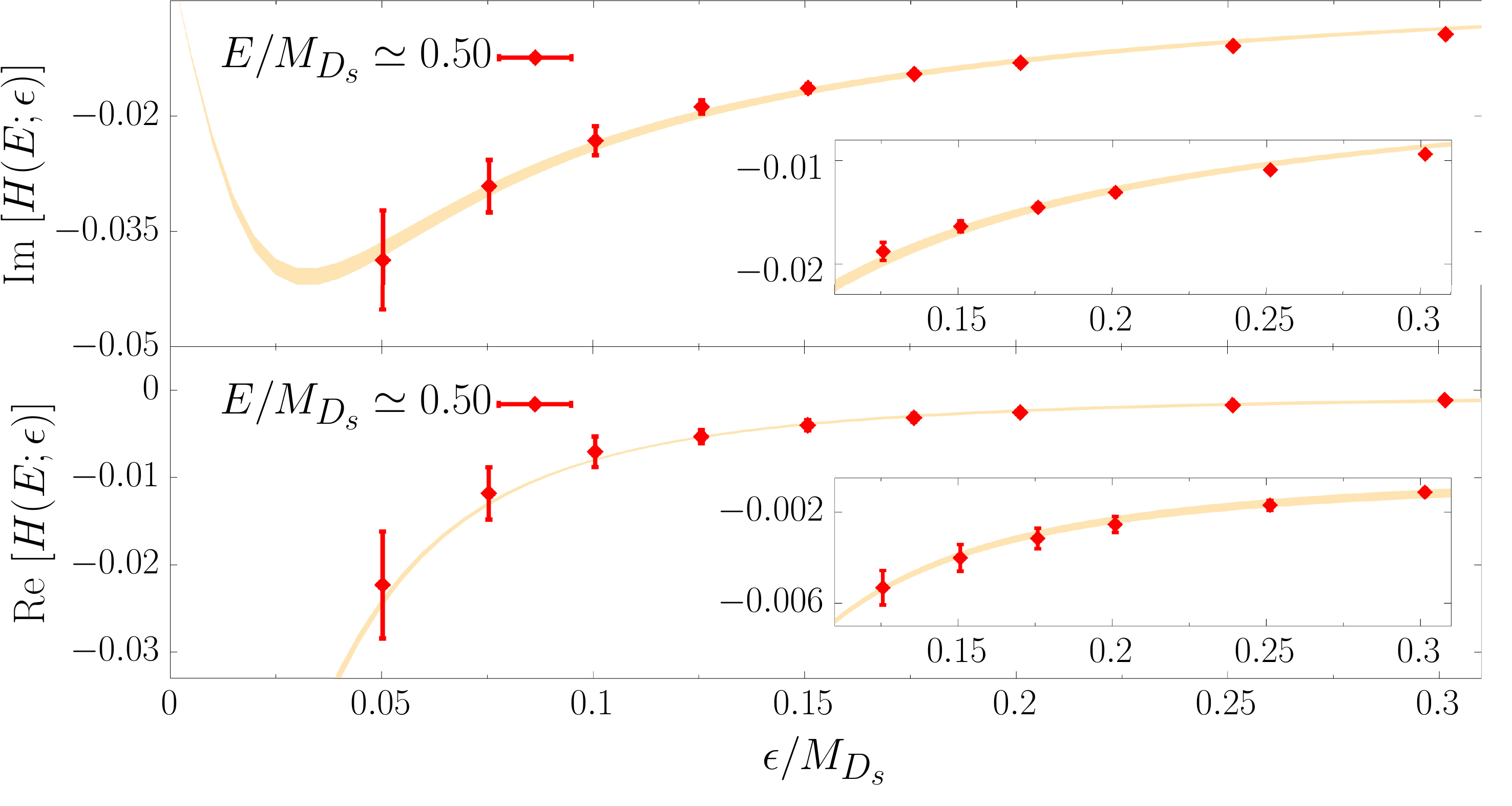} 
    \caption{\it\small The real (bottom plot) and imaginary (top plot) part of $H(E;\varepsilon)$ as a function of the smearing parameter $\varepsilon$ for a fixed value of the energy very close to the position of the $\phi$-resonance peak. The yellow-orange bands correspond to the fit performed employing the one-resonance model of Eq.~(\ref{smH-model}) (see text for details).  Results are given in lattice units.}
    \label{fig:eps_extrapolation_at_threshold}
\end{figure}

In Fig.~\ref{fig:eps_extrapolation_at_threshold} we show the dependence on the smearing parameter $\varepsilon$ of the real and imaginary part of $H(E; \varepsilon)$ for a single value of the virtual photon energy very close to the peak of the $\phi$ resonance. As the figure shows, and as expected from the previous discussion, around the threshold the $\varepsilon$ dependence is not mild. Upon decreasing $\varepsilon$ the smeared hadronic tensor keeps growing in modulus until the error become very large and $\varepsilon$ cannot be reduced further. Even though in this region we are far from the scaling regime where the $\varepsilon$-dependence is polynomial, we can still obtain useful information at the price of introducing some model dependence in the results. For instance, we can consider the one-resonance model introduced in Eq.~(\ref{eq:mod}). Around the position of the $\phi$ resonance we expect that this model provides a reasonable description of the lattice data, therefore we can fit our data for $H(E;\varepsilon)$ using the Ansatz in Eq.~(\ref{smH-model}) with $M$ and $\Gamma$ fixed to their physical values, namely $M \simeq 0.53 M_{D_{s}}$ and $\Gamma \simeq 5~{\rm MeV}$.  As Eq.~(\ref{smH-model}) shows, the introduction of a finite $\varepsilon$ in the model, simply corresponds to a shift of the width of the resonance from $\Gamma$ to $\Gamma+\varepsilon$. It is then also clear that with values of $\varepsilon$ of order $\mathcal{O}(100~{\rm MeV})$, as those used in the present analysis, a value of $\Gamma\simeq 5~{\rm MeV}$ cannot be determined from a fit of the data. In the case of the fit to ${\rm Re}[H(E;\varepsilon)]$, following the discussion in the previous section, we also introduce in the Ansatz an additional constant fit parameter whose role is to account, in an effective way, for heavier states contributions (see the discussion around Eq.~(\ref{eq:spec_decomposition})). The result of the $\chi^2-$minimization are shown by the yellow-orange bands in Fig.~\ref{fig:eps_extrapolation_at_threshold}. 
The fit yields a reduced $\chi^{2}$ smaller than unit, and effectively reproduces the $\varepsilon$-behavior of the lattice data.

\begin{figure}
    \centering
    \includegraphics[scale=0.51]{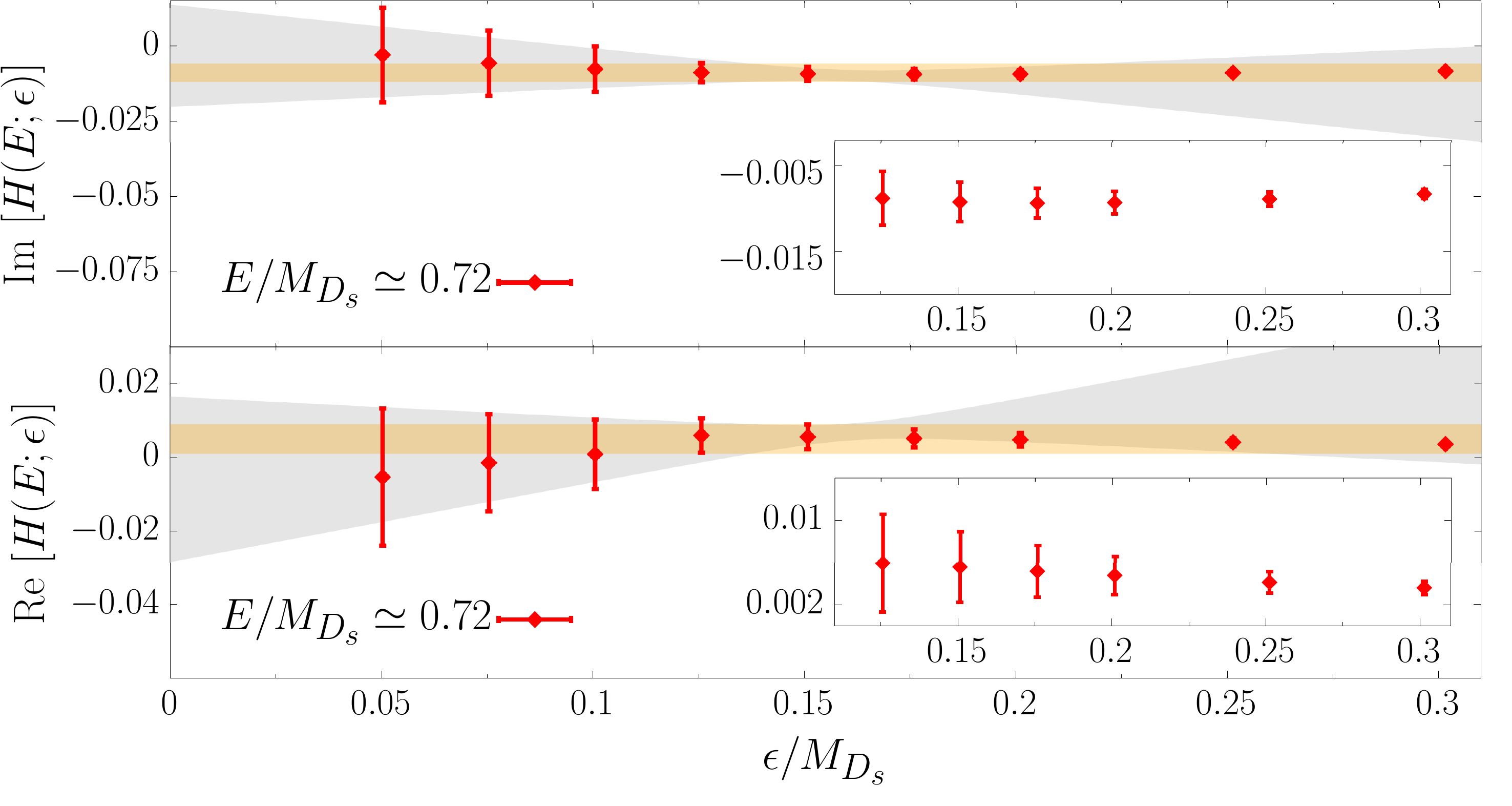} 
    \caption{\it\small The real (bottom plot) and imaginary (top plot) part of $H(E;\varepsilon)$ as a function of the smearing parameter $\varepsilon$ for a value of the energy above the position of the $\phi$-resonance peak, for which $\Delta(E) \simeq 0.2 M_{D_{s}}$. The yellow-orange and grey bands correspond respectively to the results of a constant and linear fit in the region $\epsilon \lesssim 0.15 M_{D_{s}}$. Results are given in lattice units.}
    \label{fig:eps_extrapolation_above_threshold}
\end{figure}

Away from the $\phi$-resonance peak the situation is different. This case is illustrated in Fig.~\ref{fig:eps_extrapolation_above_threshold} where we show the $\varepsilon$-dependence for a virtual photon energy $E \simeq 0.72 M_{D_{s}}$, which is sufficiently far from the $\phi$-resonance peak that the condition $\varepsilon\ll \Delta(E) \simeq 0.2 M_{D_{s}}$ is fulfilled for some of the simulated values of the smearing parameter. As Fig.~\ref{fig:eps_extrapolation_above_threshold} shows, given the relatively large statistical uncertainty of the present determination, we observe practically no dependence on $\varepsilon$ within the errors. For these value of energy the extrapolation can be trusted, and in Fig.~\ref{fig:eps_extrapolation_above_threshold} we show, the results of a constant and linear extrapolation in the region $\varepsilon/M_{D_{s}} \lesssim 0.15$, which are indicated in the figure by the yellow-orange and grey bands, respectively.  In the following, in order to be conservative, we will use the results obtained in the linear extrapolation as our final determination. 

We now present the results of the extrapolation of the smeared hadronic amplitude to vanishing $\varepsilon$. As we demonstrated and discussed earlier, for energies $E < E^{*}$, the smearing parameter $\varepsilon$ can be reduced as desired, and the results of the spectral method converge to those of the standard approach of Eq.~(\ref{ampE}) in the limit $\varepsilon \to 0$. Instead, above the threshold, where only the spectral method can be used, the current statistical accuracy of our lattice correlator $C_{E}(t)$ limits the minimum amount of smearing we can reach to order $\mathcal{O}(100~{\rm MeV})$. Therefore, within an energy region of few hundreds MeV around the resonance, a polynomial extrapolation is currently not possible, and in this region we rely on the Breit-Wigner model of Eq.~(\ref{H-model}) 
to perform the $\varepsilon\to 0$ extrapolation, as discussed above in the text.

A polynomial, model-independent, extrapolation of the data, is instead possible for energies that are sufficiently far from sharp resonances. For this reason, we performed a linear extrapolation of our data for ${\rm Re}[H(E;\varepsilon)]$ and ${\rm Im}[H(E;\varepsilon)]$ only for photon energies $E \gtrsim 1.2 M \simeq 0.65 M_{D_{S}}$, employing in the fit for each energy, only those values of the smearing satisfying $\varepsilon \lesssim 0.7 \times \Delta(E)$, where $\Delta(E)/M_{D_{s}} \simeq | E - M |/M_{D_{s}} \simeq | E/M_{D_{s}} - 0.53 |$, neglecting the very small width $\Gamma \simeq 5~{\rm MeV}$ of the $\phi$ resonance.

\begin{figure}
    \centering
    \includegraphics[scale=0.51]{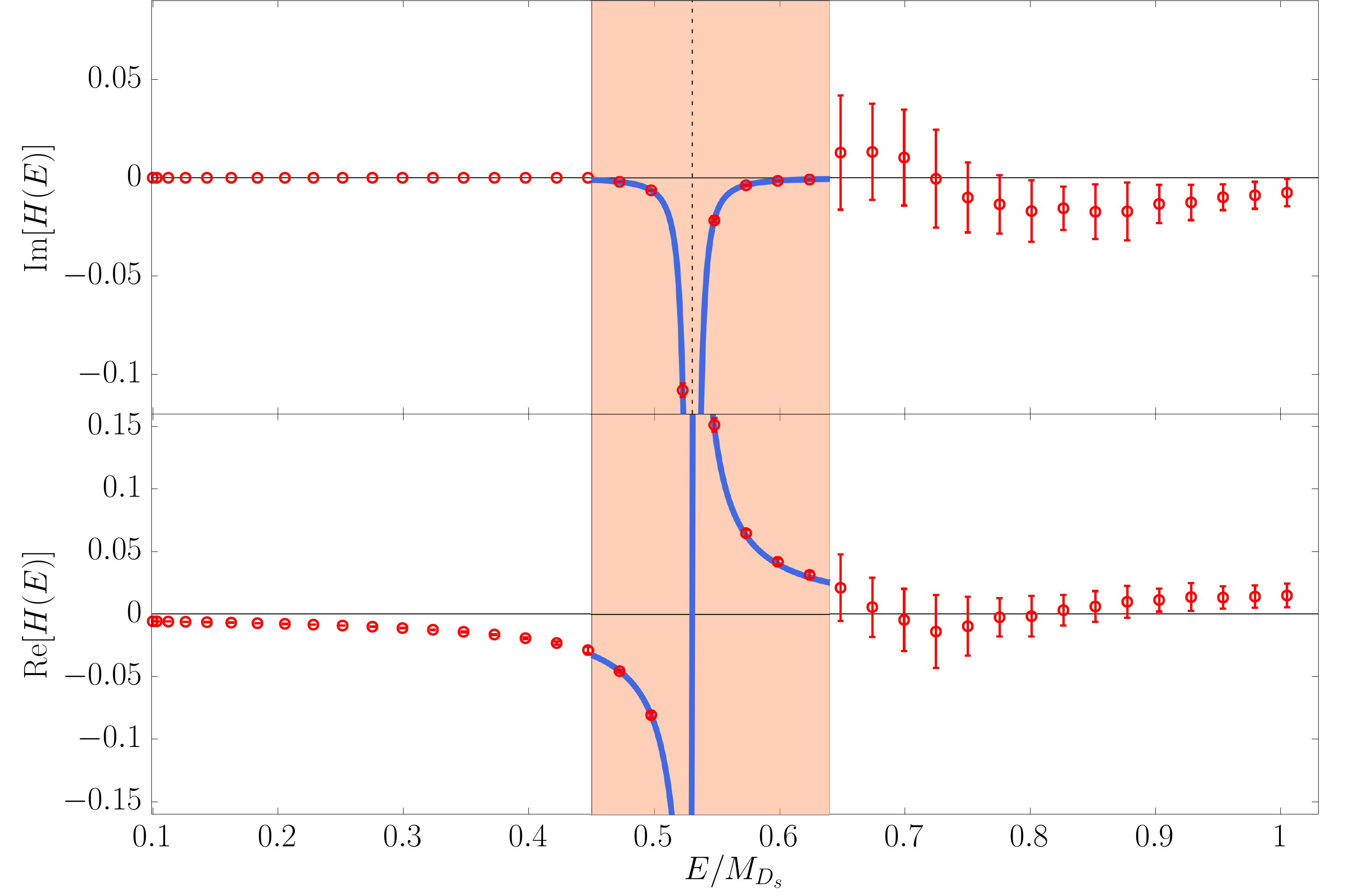} 
    \caption{\it\small Results of the extrapolation $\varepsilon\to 0$ of the real (bottom plot) and imaginary (top plot) part of $H(E;\varepsilon)$ (red data points). The orange band indicates the region where the extrapolation to vanishing $\varepsilon$ has been performed assuming that the $\varepsilon$-dependence of the lattice data is described  by the Breit-Wigner model in Eq.~(\ref{smH-model}), with $M= 0.53M_{D_{s}}$ and $\Gamma = \Gamma_{\phi}\simeq 5~{\rm MeV}$ (see Fig.~\ref{fig:eps_extrapolation_at_threshold}). The blue curves, which are given to guide the eyes, correspond to the prediction of the Breit-Wigner model of Eq.~(\ref{H-model}), in which the amplitude $A$ has been determined from the fit of the lattice data. Results are given in lattice units. }
    \label{fig:extrapolated_data}
\end{figure}

The results of the extrapolations to vanishing $\varepsilon$ are shown in Fig.~\ref{fig:extrapolated_data}. 
In the figure, the orange band represents the area around the $\phi$ resonance where no polynomial extrapolation has been attempted. In this region we perform instead the extrapolation assuming that the $\varepsilon$-dependence of the data is described by the one-resonance Breit-Wigner model in Eq.~(\ref{smH-model})
with $\Gamma = \Gamma_{\phi} \simeq 5~{\rm MeV}, M\simeq 0.53 M_{D_{s}}$ (see Fig.~\ref{fig:eps_extrapolation_at_threshold} and the corresponding text for additional details regarding the extrapolation).
As shown in the figure, the results of the extrapolation obtained using the Breit-Wigner model smoothly connect with the ones obtained through a polynomial extrapolation, at the border of the region indicated by the orange band. For $E > 0.65 M_{D_{s}}$, our results still have significant errors; however, we stress that this is a fairly low-statistics calculation of $C_{E}(t)$, and the uncertainties of the spectral method we propose are systematically improvable by computing the Euclidean correlator with increasing precision.

\section{Conclusions}

We have presented a novel method to determine, by means of Euclidean lattice calculations, the real and imaginary parts of complex electroweak amplitudes involving two external currents and a single hadron or the QCD vacuum in the external states. In this situation a direct analytic continuation of the relevant time-dependent correlation functions from Minkowskian to Euclidean spacetime is not possible, and this hinders a straightforward application of the traditional lattice techniques employed to evaluate $H(E)$. The method we propose bypasses the obstacle posed by the appearance of non-analiticities, by rewriting the hadronic amplitude $H(E)$ as a convolution integral involving the underlying spectral density $\rho(E)$ (see Eq.~(\ref{master})). As well known, the latter is related to the Euclidean correlator $C_{E}(t)$, our lattice input, through an inverse Laplace transform. The convolution integral in Eq.~(\ref{master}) that defines the hadronic amplitude $H(E)$ in terms of the spectral density, develops singularities in correspondence of any of the energies of the intermediate states, and we devise to regularize the singularities employing the Feynmann $i\varepsilon$ prescription, in the same spirit of other recent proposals built on the same technique~\cite{Bulava:2019kbi}. We have shown analytically that the introduction of a finite regulator $\varepsilon$ produces a smearing of the physical hadronic amplitude $H(E)$ over an energy radius $\varepsilon$. The resulting smeared amplitude at non-zero $\varepsilon$, which we indicated in the text with $H(E;\varepsilon)$, can be then evaluated using the HLT method developed in Ref.~\cite{Hansen:2019idp}. The method allows to determine, from the knowledge of the Euclidean correlator $C_{E}(t)$ only and with controlled errors, the convolution between spectral densities and smooth, non-singular, kernel functions. 
 
We have studied in detail the issue of the $\varepsilon \to 0$ extrapolation of the smeared amplitude $H(E;\varepsilon)$, which needs to be performed in order to recover the physical amplitude $H(E)$. We have found that in order to be able to perform a controlled $\varepsilon\to 0$ extrapolation, two conditions must simultaneously be met: on the one hand side it is required that on a finite lattice of linear extent $L$ the simulated values of $\varepsilon$ must be much larger than the typical separation between the discrete energy levels $E_{n}$ of the corresponding finite-volume Hamiltonian. This is parametrically expressed by the condition $\varepsilon L \gg 1$. On the other hand side, in order to ensure a smooth behaviour in $\varepsilon$, it is necessary that for any fixed value of the energy $E$ considered, the typical size of the interval around $E$ in which $H(E)$ is significantly varying must be smaller than $\varepsilon$. This is expressed by the condition $\varepsilon\ll \Delta(E)$, where $\Delta(E)$ is the logarithmic derivative of the hadronic amplitude $H(E)$. We have verified this explicitly in an exactly-solvable model with the spectral density given by a single Breit-Wigner resonance of mass $M$ and decay width $\Gamma$. In this model, the criterion for a smooth $\varepsilon-$convergence assumes the simple form $\varepsilon \ll \Delta(E) \equiv \sqrt{ (E-M)^2 + \Gamma^{2}}$. In a nutshell, the criterion states that the extrapolation to vanishing $\varepsilon$ is expected to be smooth away from the region where sharp resonances are present in the spectral density $\rho(E)$.  

To study the effectiveness of the spectral method, we have considered its application to the calculation of the relevant matrix elements describing the radiative leptonic decay $P\to \ell \nu \gamma^{*}$, where $\ell$ is a charged lepton, $\gamma^{*}$ a virtual photon, and in our case $P=D_{s}$. This decay channel is challenging as it develops the non-analiticity problem when the offshellness $\sqrt{k^{2}} \equiv \sqrt{ E^{2} - |\bs{k}|^{2}}$ of the virtual photon $\gamma^{*}$ exceeds the threshold for vector-meson production. In our case, this occurs in correspondence of the production of the $\phi$ resonance, i.e. when $\sqrt{k^{2}} \simeq M_{\phi}$. We have applied the spectral method, computing the relevant Euclidean correlation functions on a single gauge ensemble produced by the ETMC with $N_{f}=2+1+1$ Wilson-clover twisted-mass fermions. The HLT method has been used to reconstruct the smeared amplitudes for arbitrary photon energies $E$, and for three fixed values of the photon momentum $|\bs{k}| \simeq 0.2,0.5,0.7~{\rm GeV}$, as measured in the $D_{s}$ meson rest frame. The present statistical accuracy of the Euclidean correlator allowed us to reconstruct the smeared amplitudes $H(E;\varepsilon)$ for values of the smearing parameter $\varepsilon \gtrsim 100~{\rm MeV}$. The real and imaginary parts of the smeared amplitudes have been compared with the VMD predictions, finding overall a good qualitative agreement. The differences we observed can be attributed to the contributions of heavier states, which are not captured by the VMD model. 

As for the $\varepsilon$ extrapolation, we showed that for photon energies below the threshold we are able to recover, through the polynomial extrapolation, the result of the standard approach. 
As expected, the extrapolation turned out to be more involved around the position of the sharp $\phi$ resonance. In this case, we have found numerically that, in agreement with our theoretical analysis, the simulated values of $\varepsilon$ are far away from the scaling region where the corrections to the $\varepsilon=0$ limit can be described by a polynomial of low degree in $\varepsilon$. In this energy region, however, useful physical information can still be extracted from the data, employing a model of the hadronic amplitude $H(E)$ with parameters that can be determined through a fit to the lattice data for $H(E;\varepsilon)$. 
For photon energies sufficiently above the position of the $\phi$-resonance peak, instead, the lattice data did not show a significant $\varepsilon$-dependence, and we carried out a linear $\varepsilon$-extrapolation. In this energy region the relative statistical uncertainties are large. However, we stress that this is only a proof-of-principle calculation, and that the errors of the spectral method we propose can be systematically improved by evaluating the relevant Euclidean correlation functions with higher statistical accuracy.

In the future, we plan to extend this calculation to other pseudoscalar mesons $P$, and perform a reliable continuum limit extrapolation. In light of our findings, the ideal channels to target with the spectral method are the kaon decays $K\to \ell \nu \gamma^{*}$, which have been studied in Ref.~\cite{Gagliardi:2022szw} with unphysical pion masses. Indeed, in this case no sharp resonances are present in the spectral density, which is dominated by the contribution from the broad $\rho$ resonance. This will give us more control on the $\varepsilon\to 0$ extrapolation, possibly allowing us to determine the hadronic amplitude over the whole phase space.

.

\section{Acknowledgements}
We thank G.\,Martinelli, F.\,Mazzetti and C.T.\,Sachrajda for many useful discussions, and all the members of the ETMC for the most enjoyable collaboration. We acknowledge CINECA for the provision of CPU time under the specific
initiative INFN-LQCD123 and IscrB\_S-EPIC. F.S. G.G and S.S. are supported by the Italian Ministry
of University and Research (MIUR) under grant PRIN20172LNEEZ. F.S. and G.G are supported by
INFN under GRANT73/CALAT. F.S. is supported by ICSC – Centro Nazionale di Ricerca in High Performance
Computing, Big Data and Quantum Computing, funded by European Union –
NextGenerationEU.

\bibliography{biblio}
\bibliographystyle{JHEP}


\end{document}